\documentclass{vldb}

\newcommand{\x}[1]{{\color{red}\bf{|}}}
\newcommand{\remove}[1]{{}\xspace}

\usepackage{amsmath,amsfonts,graphicx,xspace}
\usepackage{color,url,subfigure}
\usepackage{times}

\usepackage[noend]{algorithmic}
\usepackage[boxed]{algorithm}

\newtheorem{lemma}{Lemma}
\newtheorem{theorem}[lemma]{Theorem}

\newtheorem{example}{Example}[section]

        {\medskip}
        {\hspace*{\fill}$\Box$\par\vspace{4mm}}
        {\hspace*{\fill}$\Box$\par}
\newcommand{\stitle}[1]{\vspace{0.6ex} \noindent{\bf #1}}

\newcommand{\ie}{{i.e.}\xspace}

\newcommand{\etal}{{\em et al.}\xspace}
\newcommand{\eg}{{\em e.g.}\xspace}

\newcommand{\csec}{Section~}
\newcommand{\csecs}{Sections~}
\newcommand{\cthm}{Theorem~}

\newcommand{\cequ}{Equation~}

\newcommand{\cfig}{Figure~}
\newcommand{\cfigs}{Figures~}
\newcommand{\calg}{Algorithm~}
\newcommand{\cexp}{Example~}

\newcommand{\pr}[1]{{\rm Pr}\left[#1\right]\xspace}
\newcommand{\ep}[1]{{\rm E}\left[#1\right]\xspace}
\newcommand{\bigoh}[1]{{\rm O}\!\left(#1\right)\xspace}

\newcommand{\entities}{{\cal V}\xspace}
\newcommand{\attributes}{{\cal A}\xspace}
\newcommand{\type}{\tau\xspace}
\newcommand{\attr}{\alpha\xspace}
\newcommand{\types}{{\cal C}\xspace}
\newcommand{\edges}{{\cal E}\xspace}
\newcommand{\graph}{{\cal G}\xspace}
\newcommand{\textname}{{\sf text}\xspace}
\newcommand{\pathset}{{\cal P}\xspace}

\newcommand{\query}{{\sf q}\xspace}
\newcommand{\word}{{\sf w}\xspace}

\newcommand{\pattern}{{\sf pattern}\xspace}
\newcommand{\trees}{{\sf trees}\xspace}
\newcommand{\score}{{\sf score}\xspace}
\newcommand{\scoreapprox}{\widehat{{\sf s}}\xspace}
\newcommand{\scoreexact}{{{\sf s}}\xspace}

\newcommand{\roots}{{\rm Roots}\xspace}
\newcommand{\patns}{{\rm Patterns}\xspace}
\newcommand{\paths}{{\rm Paths}\xspace}

\newcommand{\countpat}{{\sc CountPat}\xspace}

\newcommand{\sharpp}{{\sc \#P}\xspace}
\newcommand{\np}{{\sc NP}\xspace}

\newcommand{\wiki}{Wiki\xspace}
\newcommand{\imdb}{IMDB\xspace}
\newcommand{\algBaseline}{{\sf Baseline}\xspace}
\newcommand{\algLE}{{\sf LETopK}\xspace}
\newcommand{\algPE}{{\sf PETopK}\xspace}

\newcommand{\algPatternEnum}{{\sc PatternEnum}\xspace}
\newcommand{\algLinearEnum}{{\sc LinearEnum}\xspace}
\newcommand{\algLinearEnumTopK}{{\sc LinearEnum-TopK}\xspace}

\begin{document}


\title{Finding Patterns in a Knowledge Base using Keywords to Compose Table Answers}



%
%
%
%

\numberofauthors{1} 

\author{
%
%
\alignauthor
Mohan Yang$^1$
         \titlenote{Work done while visiting Microsoft Research}
~~~~~~ Bolin Ding$^2$ ~~~~~~ Surajit Chaudhuri$^2$ ~~~~~~ Kaushik Chakrabarti$^2$
         \affaddr{$^1$University of California, Los Angeles, CA}\\
         \affaddr{$^2$Microsoft Research, Redmond, WA}\\
         \email{yang@cs.ucla.edu, ~~~ {\large\{}bolind, surajitc, kaushik{\large\}}@microsoft.com
}
}

\maketitle

\begin{abstract}
We aim to provide table answers to keyword queries using a knowledge base. For queries referring to multiple entities, like ``Washington cities population'' and ``Mel Gibson movies'', it is better to represent each relevant answer as a table which aggregates a set of entities or joins of entities within the same table scheme or {\em pattern}.
%
%
%
In this paper, we study how to find highly relevant patterns in a knowledge base for user-given keyword queries to compose table answers. A knowledge base is modeled as a directed graph called knowledge graph, where nodes represent its entities and edges represent the relationships among them. Each node/edge is labeled with type and text. A pattern is an aggregation of subtrees which contain all keywords in the texts and have the same structure and types on node/edges. We propose efficient algorithms to find patterns that are relevant to the query for a class of scoring functions. We show the hardness of the problem in theory, and propose path-based indexes that are affordable in memory. Two query-processing algorithms are proposed: one is fast in practice for small queries (with small numbers of patterns as answers) by utilizing the indexes; and the other one is better in theory, with running time linear in the sizes of indexes and answers, which can handle large queries better.
%
%
We also conduct extensive experimental study to compare our approaches with a naive adaption of known techniques.
%
\end{abstract}

\section{Introduction}

Users often look for information about sets of entities, \eg, in the form of tables \cite{pvldb:LimayeSC10, pvldb:VenetisHMPSWMW11, pvldb:PimplikarS12}. For example, an analyst wants a list of companies that produces database software along with their annual revenues for the purpose of market research. Or a student wants a list of universities in a particular county along with their enrollment numbers, tuition fees and financial endowment in order to choose which universities to seek admission in.

To provide such services, some works leverage the vast corpus of HTML tables available on the Web, trying to interpret them, and return relevant ones in response to keyword queries \cite{pvldb:LimayeSC10, pvldb:VenetisHMPSWMW11, pvldb:PimplikarS12, sigmod:YakoutGCC12}. There are also two such commercial table search engines: Google Tables \cite{url:googletables} and Microsoft's Excel PowerQuery \cite{url:powerquery}.
Our work is complementary to this line, and aims to compose tables in response to keyword queries from patterns in {\em knowledge bases} when the desired tables are not available or of low quality in the corpus.

%
There are abundant sources of high-quality structured data, called {knowledge bases}: DBPedia \cite{url:dbpedia}, Freebase \cite{url:freebase}, and Yago \cite{url:yago} are examples of knowledge bases containing information on general topics, while there are also specialized ones like IMDB \cite{url:imdb} and DBLP \cite{url:dblp}.
%
%
%
A knowledge base contains information about individual {\em entities} together with {\em attributes} representing relationships among them. We can model a knowledge base as a directed graph, called {\em knowledge graph}, with {\em nodes} representing {\em entities} of different {\em types} and {\em edges} representing relationships, \ie, {\em attributes}, among entities.

We can find the subtrees of the knowledge graph that contain all the keywords and return them in ranked order (refer to Yu \etal \cite{survey:2010Yu} and Liu \etal \cite{www:LiuC11} for comprehensive surveys, and \csec\ref{sec:related} for detailed discussion). However, it is not adequate when the user's query is to look for a table of entities. As has been noticed in \cite{pvldb:WuYSIY13}, the returned subtrees with a heterogeneous mass of shapes might correspond to different interpretations of the query, and the subtrees corresponding to certain desired interpretation may not appear contiguously in the ranked order.
If the user wants to explore all subtrees of the desired interpretation, she has to examine all the returned subtrees and manually gather those corresponding to the interpretation. This is extremely labor intensive. So we propose to automatically aggregate the subtrees that contain all the keywords into distinct interpretations and produce a ranked list of such aggregations. Structural pattern of a subtree together with the mapping from the keywords to its nodes/edges represents an interpretation of the query, called {\em tree pattern}. We aggregate the subtrees based on tree patterns. Our work sharply contrasts earlier works on ranking subtrees. To the best of our knowledge, this is the first work on finding aggregations of subtrees on graphs for keyword queries.

%
%

In this paper, we propose and study the problem of finding relevant aggregations of subtrees in the knowledge graph for a given keyword query. Each answer to the keyword query is a set of subtrees -- each subtree containing all keywords and satisfying the same tree pattern.
%
%
Such an aggregation of subtrees can be output as a table of entity joins, where each row corresponds to a subtree. When there are multiple possible tree patterns, they are enumerated and ranked by their relevance to the query.

\remove{
\begin{figure*}[t]
\centering
\scriptsize
\subfigure[Entity ``SQL Server'']
{
\label{tab:infoboxes1}
\begin{tabular}{|ll|}
\hline
{SQL Server} & ({\bf Software}) 
\\ \hline
Developer: & Microsoft
\\
Genre: & Relational Database
\\
Written in: & C++
\\
\ldots: & \ldots
\\ \hline
\end{tabular}
}
~~
\subfigure[Entity ``Microsoft'']
{
\label{tab:infoboxes2}
\begin{tabular}{|ll|}
\hline
{Microsoft} & ({\bf Company}) 
\\ \hline
Founder: & Bill Gates, Paul Allen
\\
Products: & Windows, Bing, \ldots
\\
Revenue: & US\$ 77 billion
\\
\ldots: & \ldots
\\ \hline
\end{tabular}
}
~~
\subfigure[Entity ``Bill Gates'']
{
\label{tab:infoboxes3}
\begin{tabular}{|ll|}
\hline
{Bill Gates} & ({\bf Person}) 
\\ \hline
Alma mater: & Harvard University
\\
Residence: & Medina, WA, US
\\
Spouse: & Melinda Gates
\\
\ldots: & \ldots
\\ \hline
\end{tabular}
}
\caption{Examples of entities in a knowledge base}
\label{fig:infoboxes}
\end{figure*}
}

\begin{figure*}[t]
\centering
\begin{tabular}[c]{p{2in}p{6in}}
\subfigure[Entity ``SQL Server'']
{\scriptsize
\label{tab:infoboxes1}
\begin{tabular}{|ll|}
\hline
{SQL Server} & ({\bf Software}) 
\\ \hline
Developer: & Microsoft
\\
Genre: & Relational Database
\\
Written in: & C++
\\
\ldots: & \ldots
\\ \hline
\end{tabular}
}
\subfigure[Entity ``Microsoft'']
{\scriptsize
\label{tab:infoboxes2}
\begin{tabular}{|ll|}
\hline
{Microsoft} & ({\bf Company}) 
\\ \hline
Founder: & Bill Gates, Paul Allen
\\
Products: & Windows, Bing, \ldots
\\
Revenue: & US\$ 77 billion
\\
\ldots: & \ldots
\\ \hline
\end{tabular}
}
\subfigure[Entity ``Bill Gates'']
{\scriptsize
\label{tab:infoboxes3}
\begin{tabular}{|ll|}
\hline
{Bill Gates} & ({\bf Person}) 
\\ \hline
Alma mater: & Harvard University
\\
Residence: & Medina, WA, US
\\
Spouse: & Melinda Gates
\\
\ldots: & \ldots
\\ \hline
\end{tabular}
}
&
\subfigure[{\label{fig:kg} Part of a knowledge graph derived from the knowledge base in (a)-(c), and subtrees ($T_1$-$T_3$) matching to query ``database software company revenue''}]
{
\includegraphics[width=4.0in]{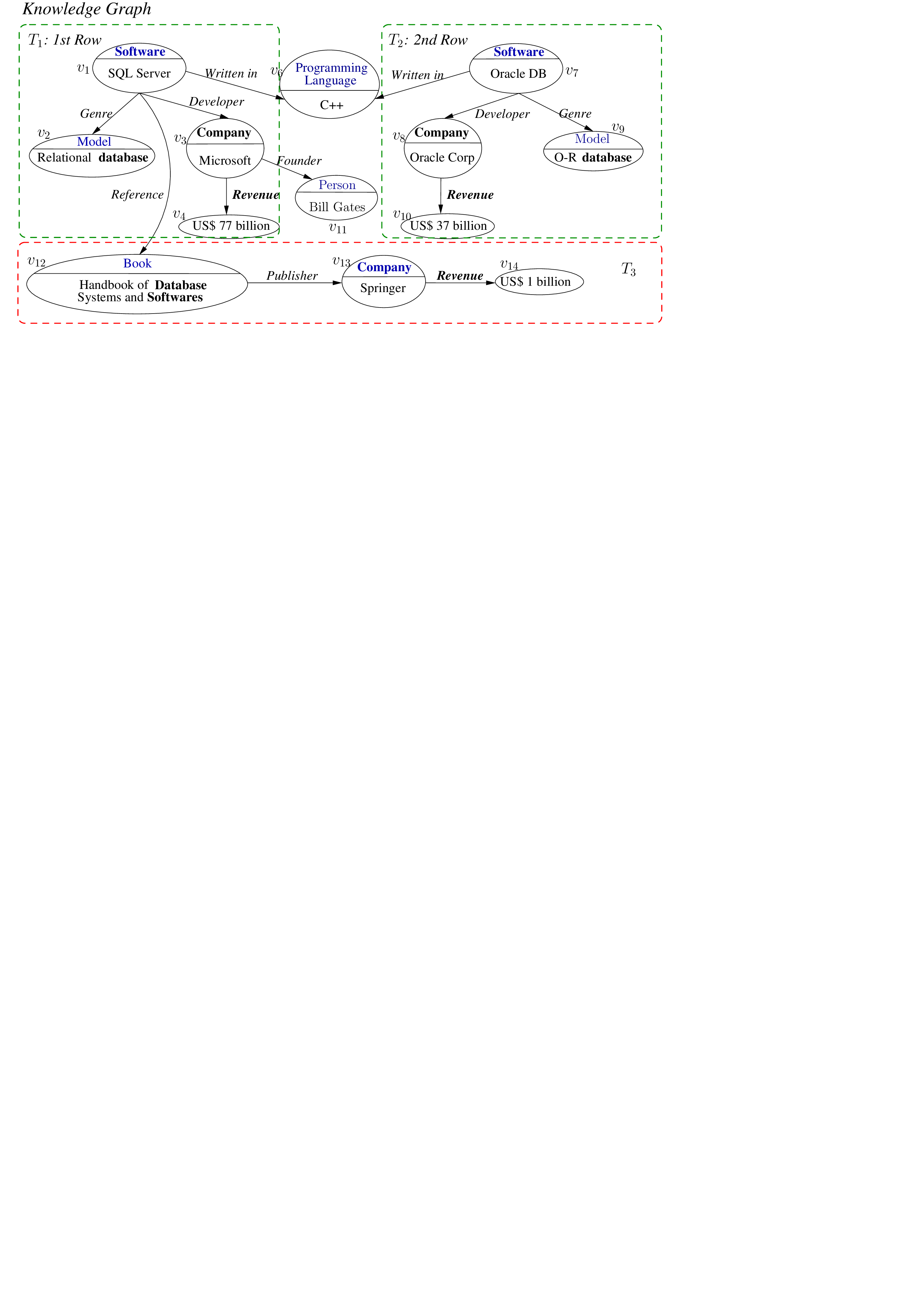}
}
\end{tabular}
\vspace{-1.6em}
\caption{(a)-(c) Entities/Attributes in Knowledge Base, (d) Knowledge Graph, Query, and Subtrees}
\vspace{-1em}
\label{fig:intro}
\end{figure*}

\begin{figure}[t]
\centering
\subfigure[Tree pattern $P_1$]
{
\label{fig:treepattern1}
\includegraphics[scale=0.43]{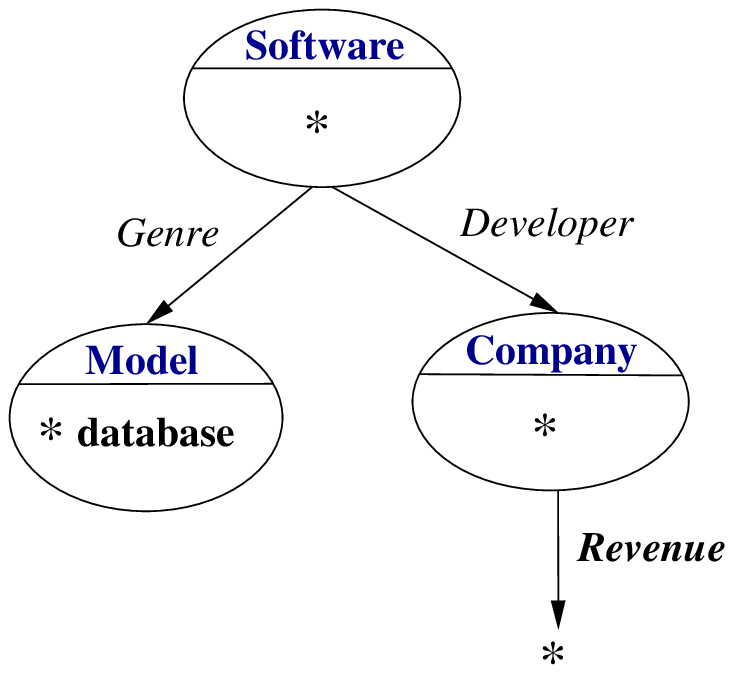}
}
~~
\subfigure[Tree pattern $P_2$]
{
\label{fig:treepattern2}
~~~~~~\includegraphics[scale=0.43]{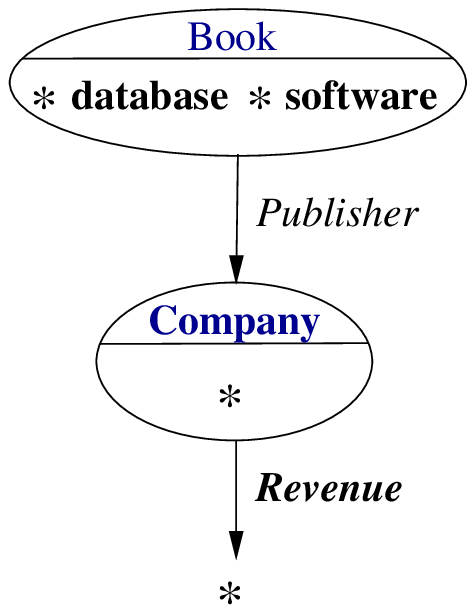}
}
\vspace{-1.6em}
\caption{Tree patterns for (a) $\{T_1, T_2\}$ and (b) $\{T_3\}$}
\vspace{-1.6em}
\label{fig:treepattern}
\end{figure}

\begin{example} \label{exa:intro} {\em (Motivation Example)}
\cfig\ref{fig:intro}(a)-(c) is a small piece of a knowledge base with three entities. For each entity (\eg, ``SQL Server'', ``Microsoft'', and ``Bill Gates''), we know its type (\eg, Software, Company, and Person, respectively), and a list of attributes (left column in \cfig\ref{fig:intro}(a)-(c)) together with their values (right column). The value of an attribute may either refer to another entity, \eg, ``Developer'' of ``SQL Server'' is ``Microsoft'', or be plain text, \eg, ``Revenue'' of ``Microsoft'' is ``US\$ 77 billion''. Such a knowledge base can be extracted from the Web like infoboxes in Wikipedia \cite{url:wiki}, or from datasets like Freebase \cite{url:freebase}.

\stitle{Knowledge graph.} A knowledge base can be modeled as a direct graph and \cfig\ref{fig:kg} shows part of such a knowledge graph. Each entity corresponds to a node labeled with its type. Each attribute of an entity corresponds to a directed edge, also labeled with its attribute type, from the entity to some other entity or plain text.


\stitle{Queries, Subtrees, and Tree Patterns.} Consider a keyword query ``database software company revenue''. Three subtrees ($T_1$, $T_2$, and $T_3$) matching the keywords are shown using dashed rectangles in \cfig\ref{fig:kg}. In subtrees $T_1$ and $T_2$, ``database'' is contained in the names of the some entities; ``software'' and ``company'' match to the types' names; and ``revenue'' matches to an attribute. Also, the structures of $T_1$ and $T_2$ are identical in terms of the types of both nodes and edges, so they belong to the same pattern in \cfig\ref{fig:treepattern1}. Similarly, $T_3$ belongs to the tree pattern in \cfig\ref{fig:treepattern2}.

\stitle{Tree patterns as answers.} A tree pattern corresponds to a possible interpretation of a keyword query, by specifying the structure of subtrees as well as how the keywords are mapped to the nodes or edges. For example, the tree pattern $P_1$ in \cfig\ref{fig:treepattern1} interprets the query as: the {\em revenue} of some {\em company} which develops {\em database} {\em software}; and $P_2$ in \cfig\ref{fig:treepattern2} means: the {\em revenue} of some {\em company} which publishes books about {\em database} {\em software}.
Subtrees of the same tree pattern can be aggregated into a table as one answer to the query, where each row corresponds to a subtree.
For example, subtrees ($T_1$ and $T_2$) of the pattern in \cfig\ref{fig:treepattern1} can be assembled into the table (the first and second rows) in \cfig\ref{fig:exampleAnswers}.
%
%
%
\end{example}

\remove{
\begin{figure}[t]
\begin{center}
\includegraphics[width=3.2in]{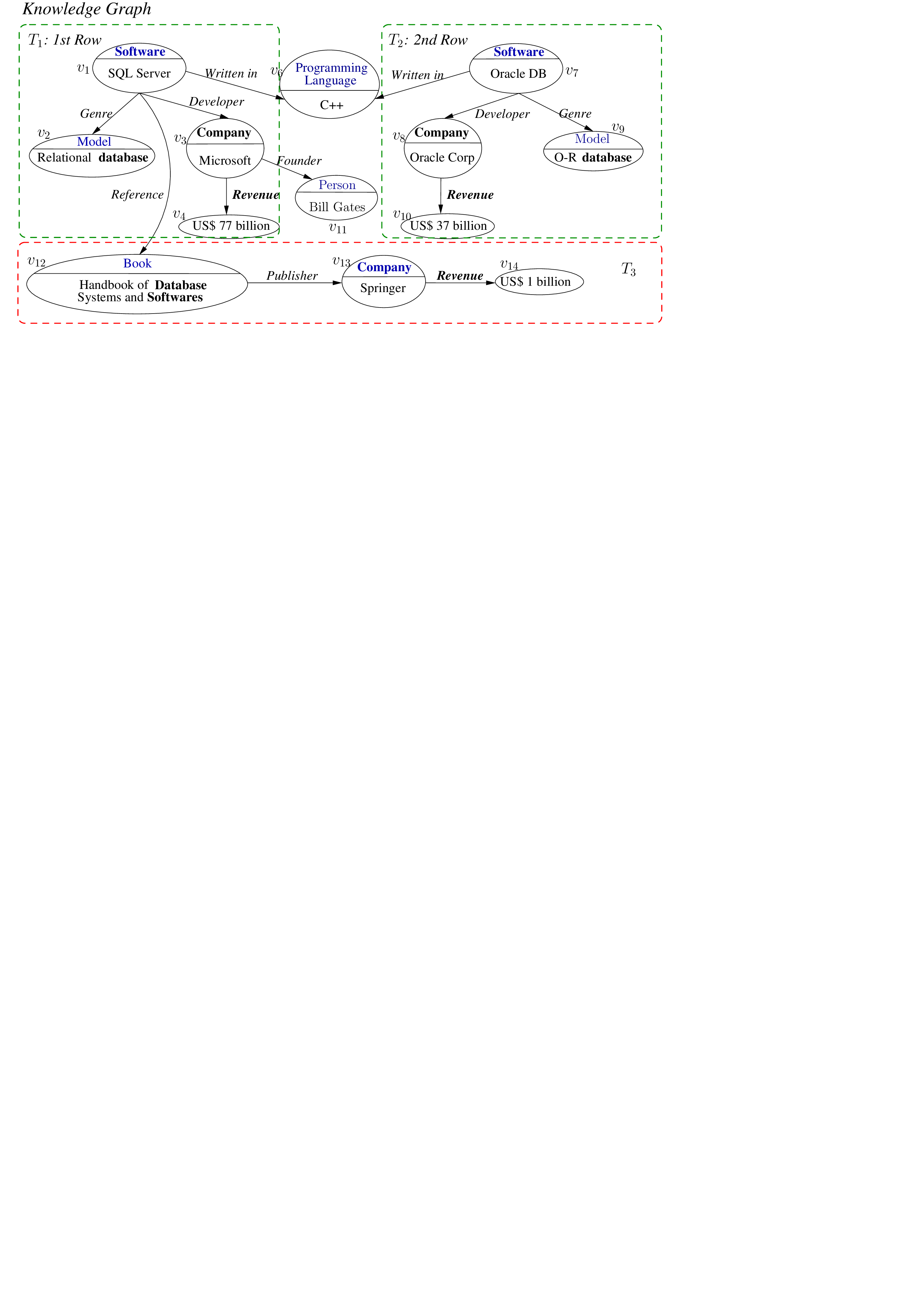}
\vspace{-0.15in}
\caption{\small \label{fig:kg} Part of a knowledge graph derived from the knowledge base in \cfig\ref{fig:infoboxes} and subtrees matching to query ``database software company revenue''}
\end{center}
\vspace{-.25in}
\end{figure}
}

%

\stitle{Contributions.}
First, we propose the problem of {\em finding relevant tree patterns in a knowledge graph}. We define {\em tree patterns} as answers to a keyword query in a knowledge graph. A class of scoring functions is introduced to measure the relevance of a pattern.

There are usually many tree patterns for a keyword query. We need efficient algorithms to enumerate these patterns and find the top-$k$. We then analyze the hardness of the problem in theory. The hardness comes from ``counting the number of paths between two nodes in the graph'', which inspires us to design two types of path-pattern based inverted indexes: paths starting from a node/edge containing some keyword and following certain pattern are aggregated and materialized in the index in memory. When processing an online query, by specifying the word and/or the path pattern, a search algorithm can retrieve the corresponding set of paths.

Two algorithms for finding the relevant tree patterns for a keyword query are proposed based on such indexes.

The first one enumerates the combinations of root-leaf path patterns in tree patterns, retrieves paths from the index for each path pattern, and joins them together on the root node to get the set of subtrees satisfying each tree pattern. Its worst-case running time is exponential in both the index size and the output size: when there are $m$ keywords and each has $p$ path patterns in the index, we need to check all the $p^m$ combinations in the worst case; but it is possible that there is no subtree satisfying any of these tree patterns. Although join operations are wasted on such ``empty patterns'', the advantage of this algorithm is that no online aggregation is required, as all subtrees with the same tree pattern are generated at one time. So it performs well in practice most of the time. 

The second algorithm tries to avoid unnecessary join operations by first identifying all {\em candidate roots} with the help of path indexes. Each candidate root reaches every keyword through at least one path pattern, so there must be some tree pattern containing a subtree with this root. Those subtrees are enumerated and aggregated for each candidate root. The running time of this algorithm can be shown to be linear in the index size and the output size. To further speed it up, we can sample a random subset of candidate roots (\eg, 10\% of them), and obtain an estimated score for each pattern based on them. Only for the patterns with the highest top-$k$ estimated scores, we retrieve the complete set of subtrees, and compute the exact scores for ranking. Note that when we apply such sampling techniques, there might be errors in the top-$k$ tree patterns. But we will show that the error can be bounded in theory, and demonstrate the effectiveness of this sampling technique in experiments.

We compare our algorithms with a straightforward adaption of previous techniques on finding subtrees in database graphs (\eg, \cite{icde:BhalotiaHNCS02, icde:DingYWQZL07, sigmod:HeWYY07, sigmod:LiOFWZ08}) in experiments.
We adapt their algorithms to enumerate all subtrees each containing all keywords as the first step. The second step is to aggregate those subtrees into a ranked list of tree patterns. Note that no ranking is required for the first step so the adapted enumeration algorithm is efficient, but the bottleneck lies on the second step. 
%
%
Efforts along this line are not helpful in solving our problem because they aim to find highly relevant subtrees while we aim to find highly relevant tree patterns.


\begin{figure}[t]
\centering
\scriptsize
{
\begin{tabular}{|c|c|c|c|}
\hline
{\bf Software} & Genre & {\bf Company} & {\bf Revenue}
\\ \hline \hline
SQL Server & Relational {\bf database} & Microsoft & US\$ 77 billion
\\
Oracle DB & O-R {\bf database} & Oracle & US\$ 37 billion
\\
\ldots & \ldots & \ldots & \ldots
\\ \hline
\end{tabular}
}
\vspace{-1em}
\caption{Example of a table aggregating subtrees of the tree pattern in \cfig\ref{fig:treepattern1}}
\vspace{-1.6em}
\label{fig:exampleAnswers}
\end{figure}


\stitle{Organization.} \csec\ref{sec:model} formally defines the concept of tree patterns as answers to keyword queries, and gives the problem statement. A baseline approach and hardness result are given at the end of \csec\ref{sec:model}.  In \csec\ref{sec:algorithm:index}, we introduce the index structures inspired by the hardness result. Two search algorithms based on the proposed path indexes are introduced in \csec\ref{sec:algorithm}. Experimental results and discussions are in \csec\ref{sec:exp}, followed by the discussion of related work in \csec\ref{sec:related}, and conclusion in \csec\ref{sec:conclude}.

%

\section{Model and Problem}
\label{sec:model}

We first formally define the graph model of a knowledge base used in this work, called {\em knowledge graph}. The model itself is not new but it servers as a general platform where our techniques introduced later can be applied. We then define {\em tree patterns}, each of which is an answer to a keyword query and aggregates a set of {\em valid subtrees} in the knowledge graph. We also introduce the class of scoring functions we use to measure the relevance of a {\em tree pattern} to a query. Finally, we formally define the problem of {\em finding top-$k$ tree patterns in a knowledge base using keywords}.

\subsection{Knowledge Graph}

A {\em knowledge base} consists of a collection of {\em entities} $\entities$ and a collection of {\em attributes} $\attributes$. Each {\em entity} $v \in \entities$ has values on a subset of {\em attributes}, denoted by $\attributes(v)$, and for each attribute $A \in \attributes(v)$, we use $v.A$ to denote its value. The value $v.A$ could be either another entity or free text.
%
%
Each entity $v \in \entities$ is labeled with a {\em type} $\type(v) \in \types$, where $\types$ is the set of all types in the knowledge base. 


It is natural to model the knowledge base as a {\em knowledge graph} $\graph$, with each entity in $\entities$ as a node, and each pair $(v, u)$ as a directed edge in $\edges$ iff $v.A = u$ for some attribute $A \in \attributes(v)$. Each node $v$ is labeled by its entity type $\type(v) = C \in \types$ and each edge $e = (v, u)$ is labeled by the attribute type $A$ iff $v.A = u$, denoted by $\attr(e) = A \in \attributes$. So we denote a {\em knowledge graph} by $\graph = (\entities, \edges, \type, \attr)$ with $\type$ and $\attr$ as {\em node type}s and {\em edge type}s, respectively.
There is {\em text description} for each entity/node type $C$, entity/node $v$, and attribute/edge type $A$, denoted by $C.\textname$, $v.\textname$, and $A.\textname$, respectively. 
In the rest of this paper, w.l.o.g., we assume that the value of an entity $v$'s attribute is always an entity in $\entities$, because if $v.A$ is plain text, we can create a {\em dummy entity} with text description exactly the same as the plain text.

\begin{example} \label{exa:kg} {\em (Knowledge Graph)}
\cfig\ref{fig:kg} shows part of the knowledge graph derived from the knowledge base in \cfig\ref{fig:intro}(a)-(c). Each node is labeled with its type $\type(v)$ in the upper part, and its text description is shown in the lower part. For nodes derived from plain text, their types are omitted in the graph. Each edge $e$ is labeled with the attribute type $\attr(e)$. Note that there could be more than one entity referred in the value of an attribute, \eg, attribute ``Products'' of entity ``Microsoft''. In that case, we can create multiple edges with the same label (attribute type) ``Products'' pointing to different entities, \eg, ``Windows'' and ``Bing''.
\end{example}

\subsection{Finding d-Height Tree Patterns}

Now we are ready to define {\em tree patterns}, \ie, answers for a given keyword query $\query = \{\word_1, \word_2, \ldots, \word_m\}$ in a knowledge graph $\graph = (\entities, \edges, \type, \attr)$. Simply put, a {\em valid subtree} w.r.t. the query $\query$ is a subtree in $\graph$ containing all keywords in the text description of its node, node type, or edge type. A {\em tree pattern} aggregates a set of valid subtrees with the same i) tree structure, ii) entity types and edge types, and iii) positions where keywords are matched.
%

\subsubsection{Valid Subtrees for Keyword Queries}

We first formally define a {\em valid subtree} $(T, f)$ w.r.t. a keyword query $\query$ in a knowledge graph $\graph$. It satisfies three conditions:
\begin{itemize}
\parskip=0.1cm
\item[i)] (Tree Structure) $T$ is a directed rooted subtree of $\graph$, \ie, it has a root $r$ and there is exactly one path from $r$ to each leaf.
\item[ii)] (Keyword Mapping) There is a mapping $f: \query \rightarrow \entities(T) \cup \edges(T)$ from words in $\query$ to nodes and edges in the subtree $T$, s.t., each word $\word \in \query$ appears in the text description of a node or node type if $f(\word) \in \entities(T)$, and appears in the text description of an edge type if $f(\word) \in \edges(T)$.
\item[iii)] (Minimality) For any leaf $v \in \entities$ with edge $e_v \in \edges$ pointing to $v$, there exists $\word \in \query$ s.t. $f(\word) = v$ or $f(\word) = e_v$. 
%
\end{itemize}
Condition ii) ensures that all words appear in subtree $T$ and specifies where they appear. Condition iii) ensures that $T$ is {\em minimal} in the sense that, under the current mapping $f$ (from words to nodes or edges wherever they appear), removing any leaf node from $T$ will make it invalid.
We will also refer to a valid subtree $(T, f)$ as $T$ if the mapping $f$ is clear from the context.

\begin{example} \label{exa:validtree} {\em (Valid Subtree)}
Consider a keyword query $\query$: ``database software company revenue'' ($\word_1$-$\word_4$). $T_1$ in \cfig\ref{fig:kg} is a valid subtree w.r.t. $\query$. The associated mapping $f$ from keywords to nodes in $T_1$ is: $f(\word_1) = v_2$ (appearing in the text description of node), $f(\word_2) = v_1$ (appearing in the node type), $f(\word_3) = v_3$ (appearing in the node type), and $f(\word_4) = (v_3, v_4)$ (appearing in the attribute type). $T_1$ is minimal and attaching any edge like $(v_1, v_6)$ or $(v_3, v_{11})$ to $T_1$ will make it invalid (violating condition iii)). Similarly, $T_2$ and $T_3$ are also valid subtrees w.r.t. $\query$.
\end{example}


\subsubsection{Tree Patterns: Aggregations of Subtrees}
\label{sec:model:problem:treepattern}

Consider a valid subtree $(T,f)$ w.r.t. a keyword query $\query$ with the mapping $f: \query \rightarrow \entities(T) \cup \edges(T)$. Before defining the {\em tree pattern} of $(T,f)$ for $\query$, we first define {\em path patterns}.

\stitle{Path patterns.} For each word $\word \in \query$, if $\word$ is matched to some node $v = f(\word)$, let $T(\word)$ be the path from the root $r$ to the node $v$: $v_1 e_1 v_2 e_2 \ldots e_{l-1} v_l$, where $v_1 = r$, $v_l = v$, and $e_i$ is the edge from $v_i$ to $v_{i+1}$. The {\em path pattern} for $\word$ is the concatenation of node/edge types on the path $T(\word)$, \ie,
\vspace{-0.3em}
\[\vspace{-0.3em}\pattern(T(\word)) = \type(v_1) \attr(e_1) \type(v_2) \attr(e_2) \ldots \attr(e_{l-1}) \type(v_l),\]
from node $v_1$ to node $v_l$. Similarly, if $\word$ is matched to some edge $e = f(\word)$, then the path pattern
\vspace{-0.3em}
\[\vspace{-0.3em}\!\!\!\!\!\!\!\!\!\!\!\!\!\!\!\!\!\!\!\!\pattern(T(\word)) = \type(v_1) \attr(e_1) \type(v_2) \attr(e_2) \ldots \attr(e_{l})\]
is the concatenation of node/edge types on the path $T(\word)$ from node $v_1 = r$ to edge $e_l = e$. The {\em length} of a path pattern, denoted by $|\pattern(T(\word))|$, is the number of nodes on path $T(\word)$.

\stitle{Tree patterns.} The {\em tree pattern} of a valid subtree $T$ w.r.t. $\query$ $=$ $\{\word_1,$ $\word_2,$ $\ldots,$ $\word_m\}$ is a vector with the $i^{{th}}$ entry as the path pattern of the root-leaf path containing the $i^{th}$ keyword $\word_i$, denoted as
\begin{equation}\label{equ:treepattern}
\pattern(T) = (\pattern(T({\word_1})), \ldots, \pattern(T({\word_m}))).
\end{equation}

The {\em height} of a tree pattern, denoted by ${\cal H}(\pattern(T))$, is the max length of the path patterns, \ie, $\max_i|\pattern(T(\word_i))|$.

Valid subtrees can be considered as ordered trees. To check whether patterns of two valid subtrees $T_1$ and $T_2$ w.r.t. query $\query$ are identical, we only need to check whether the path patterns are identical, $\pattern(T_1(\word_i)) = \pattern(T_2(\word_i))$, for each word $\word_i \in \query$. This can be done in linear time, because even without precomputation, each path pattern can be obtained by retrieving the types of node/edge on the path in order from the root $v_1$ to a leaf $v_l$ or $e_l$.

Conceptually, valid subtrees can be grouped by their patterns. For a tree pattern $P$, let $\trees(P, \query)$ be the set of all valid subtrees with the same pattern $P$ w.r.t. a keyword query $\query$, \ie, $\trees(P, \query) = \{T \mid \pattern(T) = P\}$. $\trees(P, \query)$ is also written as $\trees(P)$ if the query $\query$ is clear from the context.

\begin{example} \label{exa:answer} {\em (Tree Patterns as Answers)}
Let's continue with \cexp\ref{exa:validtree}. Tree pattern $P_1 = \pattern(T_1)$ w.r.t. query $\query$ is visualized in \cfig\ref{fig:treepattern1}. In particular, for $\word_4 = $ ``Revenue'' $\in \query$, we have $T_1(\word_4) = v_1 (v_1, v_3) v_3 (v_3, v_4)$, and $\pattern(T_1(\word_4)) =$ {\em (Software) (Developer) (Company) (Revenue)}. Similarly, for word $\word_1$, we have $\pattern(T_1(\word_1))$ $=$ {\em (Software) (Genre) (Model)}, for $\word_2$, $\pattern(T_1(\word_2)) =$ {\em (Software)}, and $\pattern(T_1(\word_3)) =$ {\em (Software) (Developer) (Company)}. Combining them together, we get the tree pattern $P_1$ in \cfig\ref{fig:treepattern1}.
It is easy to see that, in \cfig\ref{fig:kg}, $T_1$ and $T_2$ have the identical tree pattern $P_1$, and the tree pattern of $T_3$ is $P_2$, which is illustrated in \cfig\ref{fig:treepattern2}.
\end{example}

\stitle{Convert tree patterns into table answers.}
%
%
Once we have the tree pattern $P$, it is not hard to convert trees in $\trees(P)$ into a {\em table answer}. For each tree $T \in \trees(P)$, create a row in the following way: for each word $\word \in \query$ and path $T(\word) = v_1 e_1 v_2 e_2 \ldots e_{l-1} v_l$, create $l$ columns with values $v_1$, $v_2$, $\ldots$, $v_l$ and column names $\type(v_1)$, $\type(v_1)\attr(e_1)\type(v_2)$, $\ldots$, and $\type(v_{l-1})\attr(e_{l-1})\type(v_l)$, respectively. From the definition of tree patterns, we know all the rows created in this way have the same set of columns and this can be put and shown in a uniform table scheme. If an edge $e_i = (v_i, v_{i+1})$ appears in more than one root-leaf path (for different words $\word$'s), only one column needs to be created with name $\type(v_{i})\attr(e_{i})\type(v_{i+1})$ and value $v_{i+1}$. \cfig\ref{fig:exampleAnswers} shows the table answer derived from tree pattern $P_1$ in \cfig\ref{fig:treepattern1}.
%
How to name and order columns in the table answers in a more user-friendly way is also an important issue, but it is out of scope of this paper and requires more user study.
The rest of this paper will focus on how to find and rank tree patterns as it is the most challenging part of our problem.

\subsubsection{Relevance Scores of Tree Patterns}

There could be numerous tree patterns w.r.t. a given keyword query $\query$, so we need to define scoring functions to measure their relevance. We will define a general class of scoring functions, {\em the higher the more relevant}, which can be handled by our algorithms introduced later. First, the relevance score of a tree pattern is an aggregation of relevance scores of valid subtrees that satisfy this pattern, \eg, sum, average, and max of scores, or count of trees. Sum of scores and count of trees prefer tree patterns with more valid subtrees, while average and max prefer tree patterns with highly relevant individual subtrees. There is no global rule on which one is better, and the choice should be made based on extensive user study/feedback, which is out of the scope of this paper. We use sum of scores in the following part, but our approaches can be also extended to other aggregation functions.
\begin{equation} \label{equ:score:general}
\score(P, \query) = \sum_{T \in \trees(P)} \score(T, \query).
\end{equation}

The relevance score $\score(T, \query)$ of an individual valid subtree w.r.t. $\query$ may depend on several factors: 1) $\score_1(T,\query)$: size of $T$, we prefer small trees that represent compact relationship; 2) $\score_2(T,\query)$: importance score of nodes in $T$, we prefer more important nodes (\eg, with higher PageRank scores) to be included in $T$; and 3) $\score_3(T,\query)$: how well the keywords match the text description in $T$. Putting them together, we have
\begin{equation}\label{equ:score:tree}
\!\!\score(T, \query) = \score_1(T,\query)^{z_1} \cdot \score_2(T,\query)^{z_2} \cdot \score_3(T,\query)^{z_3},
\end{equation}
where $z_1$, $z_2$, and $z_3$ are constants that determine the weights of factors. These constants need to be tuned in practical system through user study. 
%
%
%
For the completeness, we give examples for scoring functions $\score_1$, $\score_2$, and $\score_3$ below. But note that they can also be replaced by other functions and more can be inserted into \eqref{equ:score:tree} if needed -- our search algorithms introduced later still work.
%

To measure the size of $T$, let $z_1 = -1$ and 
\begin{equation} \label{equ:score:1}
\score_1(T,\query) = \sum_{\word \in \query}\score_1(T(\word),\word) = \sum_{\word \in \query} |T(\word)|,
\end{equation}
where $|T(\word)|$ is the number of nodes on the path $T(\word)$.

To measure how significant nodes of $T$ are, let $z_2 = 1$ and
\begin{equation} \label{equ:score:2}
\score_2(T,\query) = \sum_{\word \in \query}\score_2(T(\word),\word) = \sum_{\word \in \query} {\sf PR}(f(\word)),
\end{equation}
where ${\sf PR}(f(\word))$ is the PageRank score of the node that contains word $\word \in \query$ (or, of the node that has an out-going edge contain word $\word$, if $f(\word)$ is an edge). The PageRank score ${\sf PR}(v)$ of a node $v$ is computed using the iterative method: the initial value of ${\sf PR}(v)$ is set to ${1/|\entities|}$ for all $v \in \entities$; and in each iteration, ${\sf PR}(v)$ is updated
\begin{equation}\nonumber
{\sf PR}(v) \leftarrow {\frac{1 - a}{|\entities|}} + a \sum_{(u, v) \in \edges} {\frac{{\sf PR}(u) }{{\rm OutDegree}(u)}},
\end{equation}
where $a = 0.85$ is the damping factor. The computation ends when ${\sf PR}(v)$ changes less than $10^{-8}$ during an iteration for all $v \in \entities$.

To measure how well the keywords match the text description in $T$, let $z_3 = 1$ and
\begin{equation} \label{equ:score:3}
\score_3(T,\query) = \sum_{\word \in \query}\score_3(T(\word),\word) = \sum_{\word \in \query} {\sf sim}(\word, f(\word)),
\end{equation}
where ${\sf sim}(\word, f(\word))$ is the Jaccard similarity between $\word$ and the text description on the entity (type) or the attribute type of $f(\word)$.

\begin{example} \label{exa:score} {\em (Relevance Score)}
Comparing the two tree patterns $P_1$ and $P_2$ in \cfig\ref{fig:treepattern} w.r.t. the query $\query$ in \cexp\ref{exa:validtree}, which one is more relevant to $\query$? First, consider valid subtrees $T_1, T_2 \in \trees(P_1)$ and $T_3 \in \trees(P_2)$ in \cfig\ref{fig:kg}, $T_3$ is smaller than $T_1$ and $T_2$ -- to measure the sizes, $\score_1(T_1, \query) = \score_1(T_2, \query) = 2+1+2+3 = 8$, and $\score_1(T_3, \query) = 1+1+2+3 = 7$. Second, assuming every node has the same PageRank score $1$, we have $\score_2(T_1, \query) = \score_2(T_2, \query) = \score_2(T_3, \query) = 4$. Third, considering the similarity between keywords and text description in valid subtrees $T_1$, $T_2$, and $T_3$, we have $\score_3(T_1, \query) = \score_3(T_2, \query) = \frac{1}{2} + 1 + 1 + 1 = 3.5$ and $\score_3(T_3, \query) = \frac{1}{6} + \frac{1}{6} + 1 + 1 = 2.33$. It can be found that while the scoring function prefers smaller trees, it also prefers tree patterns with more valid subtrees and subtrees matching to keywords in text description with higher similarity. So we have $\score(P_1, \query) > \score(P_2, \query)$ with $z_1 = -1$ and $z_2 = z_3 = 1$.
\end{example}

\subsubsection{Problem Statement}
\label{sec:model:problem:def}
We now formally define the {\em $d$-height tree pattern problem} to be solved in the rest of this paper: given a keyword query $\query$ in a knowledge graph $\graph$, the {\em $d$-height tree pattern problem} is to find all tree patterns $P$, with height {\em at most} $d$, w.r.t. $\query$. Users are usually interested in the top-$k$ answers, so we focus on generating $d$-height tree patterns with the top-$k$ highest relevance scores $\score(P, \query)$'s.

We introduce the height threshold $d$ of tree patterns for considerations of both search accuracy and efficiency. First, more compact answers (\ie, patterns with lower heights or tables with smaller numbers of columns) are usually more meaningful to users. Second, as keyword search is an online service, bounded height $d$ ensures in-time response. The setting of $d$ is {\em independent} on the number of keywords in the query, as it bounds the length of path from the root to {\em each} keyword. Such thresholds also appear in earlier work, \eg, \cite{vldb:HristidisP02} as tree size constraint, and more recent work \cite{sigmod:LiOFWZ08} as radius constraint, for similar considerations. Experimental study about the impact of $d$ will be reported in \csec\ref{sec:exp:exact}. 
%

\subsection{Enumeration-Aggregation Approach and Hardness Result}
%
\label{sec:model:problem:adp}

An obvious baseline that adapts previous works on finding subtrees in RDB graph using keywords (\eg, \cite{icde:BhalotiaHNCS02, icde:DingYWQZL07, sigmod:HeWYY07, sigmod:LiOFWZ08}) for our problem is called {\em enumeration-aggregation approach}. First, in the {\em enumeration step}, individual valid subtrees of height at most $d$ are generated one by one with an adaption of the backward search algorithm in \cite{icde:BhalotiaHNCS02}. No ranking or order of the generated subtrees is required, so the adapted algorithm in this step can ensure that, with proper preprocessing, the time needed to generate the $i$-th individual valid subtree is linear to the size of this tree, which is the best we can expect for an enumeration algorithm. Second, in the {\em aggregation step}, these valid subtrees are grouped by their tree patterns. Group-by in the second step is the bottleneck of this approach, but as the tree pattern of a subtree can be efficiently computed as discussed in \csec\ref{sec:model:problem:treepattern}, we can optimize this step using an efficient in-memory dictionary from tree patterns to valid subtrees.

Carefully-designed top-$k$ search strategies in \cite{icde:BhalotiaHNCS02, icde:DingYWQZL07, sigmod:HeWYY07, sigmod:LiOFWZ08} does not help for producing top-$k$ tree patterns, because i) no matter in which order the valid subtrees are generated, a highly relevant tree pattern may appear at the end of this order (for example, it is possible that each valid subtree of the tree pattern has low relevance, but the tree pattern has a high aggregate score because there are many such subtrees); and ii) optimization for the top-$k$ incurs additional cost (our baseline described above avoids to do so).

If we know the total number of tree patterns in advance, the enumeration-aggregation approach can early terminate as soon as we collect enough number of tree patterns during the enumeration. However, the hardness result below implies that it is impossible.
%
%
\begin{theorem} \label{thm:cntpat:hardness}
{\bf (Counting Complexity)} The problem of counting the number of tree patterns with height at most $d$ for a keyword query $\query$ in a knowledge graph (\countpat) is \sharpp-Complete.
\end{theorem}
%
%
%
\sharpp-Completeness is an analogue of \np-Completeness for counting problems. Our proof uses a reduction from the \sharpp-Complete problem {$s$-$t$ {\sc Paths} \cite{siamcomp:Valiant79}. Details are in the appendix.

%
%
%

The hardness result and the reduction inspire us to precompute and index path patterns, as introduced next in \csec\ref{sec:algorithm:index}.




%


\section{Indexing Path Patterns}
\label{sec:algorithm:index}

We propose a {\em path-pattern based index}, and it will be used to design efficient search algorithms introduced later in \csec\ref{sec:algorithm}.
%

In the index, for each keyword $\word$, we materialize all paths starting from some node (root) $r$ in the knowledge graph $\graph$, following certain pattern $P$, and ending at a node or an edge containing $\word$. Recall that a word $\word$ may be contained in the text description of a node or the type of a node/edge. These paths are grouped by root $r$ and pattern $P$. Only paths with length {\em at most} $d$ need to be stored if we are considering the $d$-height tree pattern problem. Depending on the needs of algorithms (introduced in \csecs\ref{sec:algorithm:fast} and \ref{sec:algorithm:enum}), these paths are either sorted by patterns first and then roots ({\em pattern-first path index} in \cfig\ref{fig:index:pattern}), or by roots first and then patterns ({\em root-first path index} in \cfig\ref{fig:index:root}).

The {\em pattern-first path index} (\cfig\ref{fig:index:pattern}) provides the following methods to access the paths:
\begin{itemize}
\parskip=-0.1cm
\item $\patns(\word)$: get all patterns following which some root can reach some node/edge containing $\word$.
\item $\roots(\word,P)$: get all roots which reach some node/edge containing $\word$ through some path with pattern $P$.
\item $\paths(\word,P,r)$: get all paths with pattern $P$ starting at root $r$ and ending at some node/edge containing $\word$.
\end{itemize}

Similarly, the {\em root-first path index} (\cfig\ref{fig:index:root}) provides the following methods to access the paths:
\begin{itemize}
\parskip=-0.1cm
\item $\roots(\word)$: get all root nodes which can reach some node/edge containing $\word$.
\item $\patns(\word,r)$: get all patterns following which the root $r$ can reach some node/edge containing $\word$.
\item $\paths(\word,r)$: get all paths which start at root $r$ and end at some node/edge containing $\word$.
\item $\paths(\word,r,P)$: get all paths with pattern $P$ starting at root $r$ and ending at some node/edge containing $\word$.
\end{itemize}

\begin{figure}[t]
\centering
\subfigure[Pattern-first path index]{\label{fig:index:pattern}
\includegraphics[scale = 0.6]{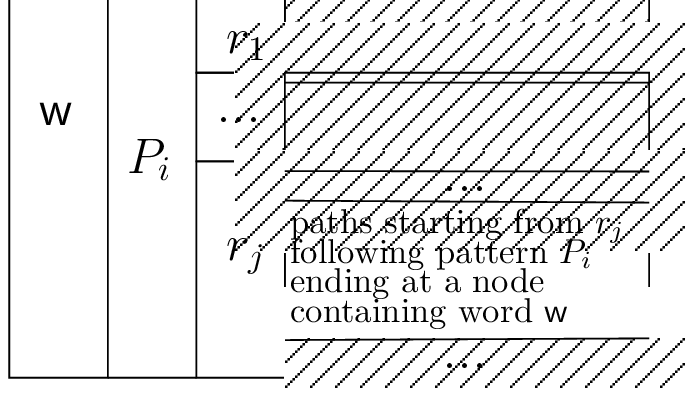}
}
\subfigure[Root-first path index]{\label{fig:index:root}
\includegraphics[scale = 0.6]{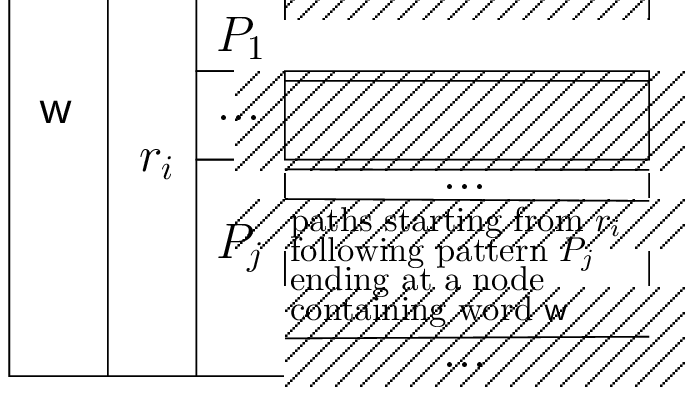}
}
\vspace{-1em}
\caption{Indexing patterns of paths ending at each word $\word$ with length no more than $d$}
\vspace{-1em}
\label{fig:index}
\end{figure}

\begin{figure}[t]
\scriptsize
\center
\subfigure[Pattern-first path index for word ``database''.\label{fig:exa:index:2}]{
\begin{tabular}{llll}
word & pattern & root & path
\\ \hline
database & (Software)(Genre)(Model) & $v_1$ & $v_1v_2$
\\ \hline
database & (Software)(Genre)(Model) & $v_7$ & $v_7 v_9$
\\ \hline
database & (Software)(Reference)(Book) & $v_1$ & $v_1v_{12}$
\\ \hline
database & (Book) & $v_{12}$ & $v_{12}$
\\ \hline
$\cdots$ & $\cdots$ & $\cdots$ & $\cdots$ 
\\ \hline
\end{tabular}
}
\subfigure[Root-first path index for word ``database''.\label{fig:exa:index:1}]{
\begin{tabular}{llll}
word & root & pattern & path
\\ \hline
database & $v_1$ & (Software)(Genre)(Model) & $v_1 v_2$
\\ \hline
database & $v_1$ & (Software)(Reference)(Book) & $v_1v_{12}$
\\ \hline
database & $v_7$ & (Software)(Genre)(Model) & $v_7v_9$
\\ \hline
database & $v_{12}$ & (Book) & $v_{12}$
\\ \hline
$\cdots$ & $\cdots$ & $\cdots$ & $\cdots$ 
\\ \hline
\end{tabular}
}
\vspace{-1em}
\caption{Examples of two types of path indexes for the knowledge graph in \cfig\ref{fig:kg}}
\vspace{-1em}
\label{fig:exa:index}
\end{figure}

Following is a tiny example of how to access these two different types of indexes.

\begin{example} \label{exa:index}
For the knowledge graph in \cfig\ref{fig:kg}, \cfig\ref{fig:exa:index} shows the two types of indexes on word $\word =$ ``database''.

For the pattern-first path index in \cfig\ref{fig:exa:index:2}, $\patns(\word)$ returns three patterns. Consider the pattern $P = $ {\em (Software) (Reference) (Book)}, $\roots(\word, P)$ returns one root $\{v_1\}$.

For the root-first path index in \cfig\ref{fig:exa:index:1}, $\roots(\word)$ returns three roots $\{v_1, v_7, v_{12}\}$. $\patns(\word, v_1)$ returns two patterns. Consider the pattern $P = $ {\em (Software) (Genre) (Model)} in particular, $\paths(\word, v_1, P)$ returns only one path $\{v_1v_2\}$.
\end{example}

\subsubsection*{Index Construction }
To construct the indexes for a (user-specified) height threshold $d$, for each possible root $r$, we use DFS to find all paths $p$ starting from $r$ and ending at some node $t$/edge $e$ with length no more than $d$. Let $\textname(p)$ be the set of words in the text description or type of the node $t$/edge $e$, and recall $\pattern(p)$ is the path pattern of $p$. The index construction process is illustrated in \calg\ref{alg:index}: each path $p$, together with its starting node $r$ and pattern $P$, is inserted into proper positions of the two indexes in lines~5-6 (we use ``+'' to denote the insertion of an element into a dictionary).

The same set of paths are stored in these two types of indexes, but in different orders. We can use dictionary data structures, such as hash tables, to support the access methods $\roots()$, $\patns()$, and $\paths()$ (in constant time).
But to improve the efficiency of the access methods in practice, we then sort and store paths sequentially in memory: by patterns first and then roots for {\em pattern-first path index} as in \cfig\ref{fig:index:pattern}, or by roots first and then patterns for {\em root-first path index} as in \cfig\ref{fig:index:root}.
Also, we store pointers pointing to the beginning of a list of paths with the same root $r$ and/or pattern $P$ to support the above access methods,

Note that the terms like $|T(\word)|$, ${\sf PR}(f(\word))$, and ${\sf sim}(\word, f(\word))$ in the relevance-scoring functions \eqref{equ:score:1}-\eqref{equ:score:3} can be precomputed and stored in the path index as well, so that the overall score \eqref{equ:score:general} can be computed efficiently online for a tree pattern.

\begin{algorithm}[th]
\underline{Input:} knowledge graph $\graph$ and height threshold $d$

\begin{algorithmic}[1]
\STATE For each node $r$ in the knowledge graph $\graph$
\STATE ~~~ For each path $p$ starting from $r$ with length $\leq d$
\STATE ~~~ ~~~ Let $P$ be $\pattern(p)$, the path pattern of $p$.
\STATE ~~~ ~~~ For each word $\word$ in $\textname(p)$
\STATE ~~~ ~~~ ~~~ \underline{Construct pattern-first path index}:
\\
~~~ ~~~ ~~~ $\patns(\word) \leftarrow \patns(\word) + P$;
\\
~~~ ~~~ ~~~ $\roots(\word, P) \leftarrow \roots(\word, P) + r$;
\\
~~~ ~~~ ~~~ $\paths(\word, P, r) \leftarrow \paths(\word, P, r) + p$.
\STATE ~~~ ~~~ ~~~ \underline{Construct root-first path index}:
\\
~~~ ~~~ ~~~ $\roots(\word) \leftarrow \roots(\word) + r$;
\\
~~~ ~~~ ~~~ $\patns(\word, r) \leftarrow \patns(\word, r) + P$;
\\
~~~ ~~~ ~~~ $\paths(\word, r, P) \leftarrow \paths(\word, r, P) + p$.
\\
~~~ ~~~ ~~~ ($\paths(\word, r)$ is supported by enumerating $P$\\~~~ ~~~ ~~~ and accessing $\paths(\word, r, P)$ for each $P$)
\end{algorithmic}
\caption{Constructing the two types of indexes}
\label{alg:index}
\end{algorithm}

We can show that the size of our index is bounded by the total number of paths with length at most $d$ and the size of text on entities and attributes.
As these paths can be enumerated in linear time, the time to compute our path index is linear in the total number of paths and the size of text, with a logarithmic factor for sorting.

\begin{theorem}
{\bf (Index Cost)} Let $\pathset$ be the set of paths in the index (with length at most $d$). For each $s$-$t$ path $p \in \pathset$, let $|p|$ be its length, $\textname(p)$ be the text on the node $t$, and $|\textname(p)|$ be the number of words in the text. Then both the root-first and the pattern-first path indexes need space ${\rm O}(\sum_{p \in \pathset} |p| \cdot |\textname(p)|)$, and can be constructed in linear time ${\rm O}(\log|\pathset|\sum_{p \in \pathset} |p| \cdot |\textname(p)|)$.
\end{theorem}

In practice, to handle synonyms, every word has its stemmed version and synonyms in our index pointing to the same path-pattern entry. The size of the index does not increase much.

\section{Searching with Path Index}
\label{sec:algorithm}

%
Two search algorithms for the $d$-height tree pattern problem are introduced in \csecs\ref{sec:algorithm:fast} and \ref{sec:algorithm:enum}: the first one performs well in practice but has exponential running time in the worst case; and the second one provides provable performance guarantee and can be further speedup using sampling techniques.
Both of them utilize the path-pattern based index introduced in \csec\ref{sec:algorithm:index}.



\subsection{Pattern Enumeration-Join Approach}
\label{sec:algorithm:fast}

From the definition of a tree pattern in \cequ\eqref{equ:treepattern}, we can see that it is composed of $m$ path patterns if there are $m$ keywords in the query. Our first algorithm enumerates the combinations of these $m$ path patterns in a tree pattern using the pattern-first path index (\cfig\ref{fig:index:pattern}); for each combination, retrieves paths with these patterns from the index, and joins them at the root to check whether the tree pattern is empty (\ie, whether there is any valid subtree with this pattern). For each nonempty one, the valid subtrees in $\trees(P)$ and its score are then computed using the same index.

The algorithm, named as \algPatternEnum, is described in \calg\ref{alg:patternenum}. It first enumerates the root type of a tree pattern in line~2. For each root type $C$, it then enumerates the combinations of path patterns starting from $C$ and ending at keywords $\word_i$'s in lines~4-8. Each combination of $m$ path patterns forms a tree pattern $P$, but it might be empty. So lines~5-6 check whether $\trees(P)$ is empty again using the path index in lines~7-8. For each nonempty tree pattern, its score and the valid subtrees in $\trees(P)$ are computed and inserted into the queue $Q$ in line~8. After every root type is considered, the top-$k$ $d$-height tree patterns in $Q$ can be output.

\begin{algorithm}[ht]
\underline{Input:} knowledge graph $\graph$, with pattern-first path index, and keyword query $\query = \{\word_1, \ldots, \word_m\}$
\begin{algorithmic}[1]
\STATE Initialize a queue $Q$ of tree patterns, ranked by scores.
\STATE For each type $C \in \types$
\STATE ~~~ Let $\patns_C(\word_i)$ be the set of path patterns \\ ~~~ rooted at the type $C$ in $\patns(\word_i)$.
\STATE ~~~ For each tree pattern $P = (P_1, \ldots, P_m)$ \\ ~~~ ~~~~~~~~~~~~ $\in \patns_C(\word_1) \times \ldots \times \patns_C(\word_m)$
\\     ~~~~~~ \underline{Check whether $\trees(P)$ is empty:}
\STATE ~~~~~~ Compute candidate roots $R \leftarrow \bigcap_{i = 1}^m \roots(\word_i, P_i)$;
\STATE ~~~~~~ If $R \neq \emptyset$ then
\STATE ~~~~~~~~~ $\trees(P) \leftarrow \bigcup_{r \in R} \paths(\word_1, P_1, r)$
\\     ~~~~~~~~~ ~~~~~~~~~~~~~~~~~~~~~~ $\times \ldots \times \paths(\word_m, P_m, r)$;
\STATE ~~~~~~~~~ Compute $\score(P,\query)$ and insert $P$ into queue $Q$.
\\     ~~~~~~~~~ (only need to maintain $k$ tree patterns in $Q$)
\STATE Return the top-$k$ tree patterns in $Q$ and valid subtrees.
\end{algorithmic}
\caption{\algPatternEnum: finding top-$k$ tree patterns by enumerating all possible tree patterns for a keyword query}
\label{alg:patternenum}
\end{algorithm}

\begin{example} \label{exa:patternenum}
Consider a query ``database software company revenue'' with four keywords $\word_1$-$\word_4$ in the knowledge graph in \cfig\ref{fig:kg}. When the root type $C = $ {\em Software}, we have two path patterns {\em (Software) (Genre) (Model)} and {\em (Software) (Reference) (Book)} from $\patns_C(\word_1)$, as in \cfig\ref{fig:exa:index:2}. To form the tree pattern in \cfig\ref{fig:treepattern1}, in line~4, we pick the first path pattern from $\patns_C(\word_1)$, {\em (Software)} from $\patns_C(\word_2)$, {\em (Software) (Developer) (Company)} from $\patns_C(\word_3)$, and {\em (Software) (Developer) (Company) (Revenue)} from $\patns_C(\word_4)$. We then find this tree pattern is not empty, and paths in the index with these patterns can be joined at nodes $v_1$ and $v_7$, forming two valid subtrees $T_1$ and $T_2$, respectively, in \cfig\ref{fig:kg}.
\end{example}

In the experiments, we will show that \algPatternEnum is efficient especially for queries which have relatively small numbers of tree patterns and valid subtrees. The advantage of this algorithm is that valid subtrees with the same pattern are generated at one time, so no online aggregation is needed. The path index has materialized aggregations of paths which can be used to check whether a tree pattern is empty and to generate valid subtrees. Also, it keeps at most $k$ tree patterns and the corresponding valid subtrees in memory and thus has very small memory footprint.

However, in the worst case, its running time is still exponential both in the size of index and in the number of valid subtrees, mainly because unnecessary costly set-intersection operators are wasted on empty tree patterns (line~5). Consider such a worst-case example: In a knowledge graph, we have two nodes $r_1$ and $r_2$ with the same type $C$; $r_1$ points to $p$ nodes $v_1, \ldots, v_p$ of types $C_1, \ldots, C_p$ through edges of types $A_1, \ldots, A_p$; and $r_2$ points to another $p$ nodes $v_{p+1}, \ldots, v_{2p}$ of types $C_{p+1}, \ldots, C_{2p}$ through edges of types $A_{p+1}, \ldots, A_{2p}$. We have two words $\word_1$ and $\word_2$, $\word_1$ appearing in $v_1, \ldots, v_p$ and $\word_2$ appearing in $v_{p+1}, \ldots, v_{2p}$. To answer the query $\{\word_1, \word_2\}$, algorithm \algPatternEnum enumerates a total of $p^2$ combined tree patterns $(CA_iC_i, CA_jC_j)$'s for $i = 1, \ldots, p$ and $j = p+1, \ldots, 2p$, but they are all empty. So its running time is $\Theta(p^2)$ or $\Theta(p^m)$ in general for $m$ keywords, where $p$ is in the same order as the size of the index.

\subsection{Linear-Time Enumeration Approach}
\label{sec:algorithm:enum}

We now introduce an algorithm to enumerate tree patterns for a given keyword
%
%
using the root-first path index (\cfig\ref{fig:index:root}). This algorithm is {\em optimal for enumeration} in the sense that its running time is linear in the size of the index and linear in the size of the answers (all valid subtrees). We prove its correctness and complexity. We will also introduce how to extend it for finding the top-$k$, and how to further speed it up using sampling techniques.

The algorithm, \algLinearEnum in \calg\ref{alg:linearenum}, is based on the following idea: instead of enumerating all the tree patterns directly, we first find all possible roots for valid subtrees, and then assemble the trees from paths with these roots by looking up the path index.

These candidate roots, denoted as $R$, can be found based on the simple fact that a node in the knowledge graph is the root of some valid subtree if and only if it can reach every keyword at some node. So the set $R$ can be obtained by taking the intersection of $\roots(\word_1), \ldots, \roots(\word_m)$ from the root-first path index (line~1).

For each candidate root $r$, recall that, using the path index, we can retrieve all patterns following which $r$ can reach keyword $\word_i$ at some node by calling $\patns(\word_i, r)$. So pick any pattern $P_i \in \patns(\word_i,r)$ for each $\word_i$, $P = (P_1, \ldots, P_m)$ is a nonempty tree pattern (\ie, $\trees(P) \neq \emptyset$). Line~7 of subroutine {\sc ExpandRoot} in \calg\ref{alg:linearenum} gets all such patterns. Each $P$ must be nonempty (with at least one valid subtree), because by picking any path $p_i$ from $\paths(\word_i, r, P_i)$ for each $P_i$, we can get a valid subtree $(p_1, \ldots, p_m)$ with pattern $P$, as in line~10. Note that valid subtrees with pattern $P$ may be under different roots, so we need a dictionary, ${\rm TreeDict}$ in line~11, to maintain and aggregate the valid subtrees along the whole process. Finally, ${\rm TreeDict}[P]$ is the set of valid subtrees with pattern $P$ as returned in lines~5-6.

\begin{example}\label{exa:linearenum}
Consider a query ``database software company revenue'' with four keywords $\word_1$-$\word_4$ in the knowledge graph in \cfig\ref{fig:kg}. The candidate roots we get are $\{v_1,$ $v_7,$ $v_{12}\}$ (line~1 of \calg\ref{alg:linearenum}). For $v_1$ and $\word_1 =$ ``database'', we can get two path patterns from $\patns(\word_1, v_1)$: {\em (Software) (Genre) (Model)}, and {\em (Software) (Reference) (Book)}. Picking the first one, together with patterns {\em (Software)}, {\em (Software) (Developer) (Company)}, and {\em (Software) (Develop) (Company) (Revenue)} for the other three keywords ``software'', ``company'', `revenue'', respectively, we can get the tree pattern in \cfig\ref{fig:treepattern1} (one of ${\cal T}$ obtained in line~7). This pattern must be nonempty, because we can find a valid subtree under $v_1$ by assembling the four paths $v_1v_2$, $v_1$, $v_1v_3$, and $v_1v_3v_4$ into a subtree $T_1$ in \cfig\ref{fig:kg} (line~10).

Another valid subtree, $T_2$ in \cfig\ref{fig:kg}, with the same pattern can be found later when candidate root $v_7$ is considered. They are both maintained in the dictionary ${\rm TreeDict}$.
\end{example}

\begin{algorithm}[htbp]
\underline{Input:} knowledge graph $\graph$, root-first path indexes, and keyword query $\query = \{\word_1, \ldots, \word_m\}$
\begin{algorithmic}[1]
\STATE Compute candidate roots $R \leftarrow \bigcap_{i=1}^m \roots(\word_i)$.
\STATE Initialize a dictionary ${\rm TreeDict}[]$.
\STATE For each candidate root $r \in R$
\STATE ~~~ Call {\sc ExpandRoot}$(r, ~ {\rm TreeDict}[])$.
\STATE For each tree pattern $P$, $\trees(P) \leftarrow {\rm TreeDict}[P]$.
\STATE Return tree patterns and valid subtrees in $\trees(\cdot)$.
\\     \vspace{+0.1cm}\hspace{-0.66cm}
Subroutine {\sc ExpandRoot}$(\hbox{root} ~ r, ~ \hbox{dictionary} ~ {\rm TreeDict}[])$
\\     ~~~ \underline{Pattern Product:}
\STATE ~~~ ${\cal T} \leftarrow \patns(\word_1, r) \times \ldots \times \patns(\word_m, r)$;
\STATE ~~~ For each tree pattern $P = (P_1, \ldots, P_m) \in {\cal T}$
\\     ~~~~~~ \underline{Path Product:}
\STATE ~~~~~~ For each $(p_1, \ldots, p_m) \in $ \\ ~~~~~~~~~~~~~~~~~~ $\paths(\word_1, r, P_1) \times \ldots \times \paths(\word_m,r,P_m)$
\STATE ~~~~~~~~~ Construct tree $T$ from the $m$ paths $p_1, \ldots, p_m$;
\STATE ~~~~~~~~~ ${\rm TreeDict}[P] \leftarrow {\rm TreeDict}[P] \bigcup \{T\}$.
\end{algorithmic}
\caption{\algLinearEnum: finding tree patterns by enumerating valid subtrees rooted from each candidate root for a keyword query}
\label{alg:linearenum}
\end{algorithm}

\algLinearEnum is optimal in the worst case because it does not waste time on invalid (empty) tree patterns. Every tree pattern it tries in line~8 has at least one valid subtree. And to generate each valid subtree, the time it needs is linear in its tree size (line~10). We formally present its correctness and complexity as follows.

\begin{theorem}\label{thm:linearenum:correct}
{\bf (Running Time and Correctness)} For a keyword query $\{\word_1, \ldots, \word_m\}$ against a knowledge graph $\graph$, let $S_i$ be the size of the path index for word $\word_i$, and let $N$ be the total number of valid subtrees. \algLinearEnum can correctly enumerate all tree patterns and valid subtrees in time $\bigoh{N \cdot d \cdot m + \sum_{i=1}^m S_i}$.
\end{theorem}

\subsubsection{Partitioning by Types to Find Top-$k$}
\label{sec:algorithm:enum:topk}

Now we introduce how to extend \algLinearEnum in \calg\ref{alg:linearenum} to find the top-$k$ tree patterns (with the highest scores). A naive method is to compute the score $\score(P, \query)$ for every tree pattern after we run \algLinearEnum for the given keyword query $\query$ on the knowledge graph $\graph$. An obvious deficiency of this method is that the dictionary ${\rm TreeDict}[]$ used in \calg\ref{alg:linearenum} could be very large (may not fit in memory and may incur higher random-access cost for lookups and insertions), as it keeps every tree patterns and associated valid subtrees, but we only require the top-$k$.

A better idea is to apply \algLinearEnum for candidate roots with the same type at one time. For each type $C$, we apply \algLinearEnum only for candidate roots with type $C$ (only line~3 of \calg\ref{alg:linearenum} needs to be changed); then compute the scores of resulting tree patterns/answers but only keep the top-$k$ tree patterns; and repeat the process for another root type. In this way, the size of the dictionary ${\rm TreeDict}[]$ is upper-bounded by the number of valid subtrees with roots of the same type, which is usually much smaller than the total number of valid subtrees in the whole knowledge graph.

For example, for the knowledge graph and the keyword query in \cfig\ref{fig:kg}, the tree pattern $P_1$ in \cfig\ref{fig:treepattern1} is found and scored when we apply \algLinearEnum for the type ``Software'', and $P_2$ in \cfig\ref{fig:treepattern2} is found when ``Book'' is the root type.

This idea, together with the sampling technique introduced a bit later, will be integrated into \algLinearEnumTopK in \calg\ref{alg:linearenum:topk} for finding the top-$k$ $d$-height tree patterns.

\begin{algorithm}[htbp]
\underline{Input:} knowledge graph $\graph$, with both path indexes, and keyword query $\query = \{\word_1, \ldots, \word_m\}$
\\
\underline{Parameters:} sampling threshold $\Lambda$ and sampling rate $\rho$
\\ \vspace{-0.4cm}
\begin{algorithmic}[1]
\STATE Initialize a queue $Q$ of tree patterns, ranked by scores.
\STATE For each type $C$ among all types $\types$
\STATE ~~~ Compute candidate roots of type $C$: \\ ~~~~~~ $R = (\bigcap_{i=1}^m \roots(\word_i)) \bigcap C$;
\STATE ~~~ Compute the number of valid subtrees rooted in $R$: \\ ~~~~~~ $N_R = \sum_{r \in R} \prod_{i=1}^m |\paths(\word_i, r)|$;
\STATE ~~~ If $N_R \geq \Lambda$ let $rate = \rho$ else $rate = 1$;
\STATE ~~~ Initialize dictionary ${\rm TreeDict}[]$;
\STATE ~~~ For each candidate root $r \in R$, \\ 
\STATE ~~~~~~ \underline{With probability $rate$,} \\ ~~~~~~ call {\sc ExpandRoot}$(r, {\rm TreeDict[]})$;
\STATE ~~~ For each tree pattern $P$ rooted at $C$ in ${\rm TreeDict}$
\STATE ~~~~~~ Compute estimated score $\scoreapprox(P, \query)$ ($\approx \score(P, \query)$) \\ ~~~~~~ from sample valid subtrees in ${\rm TreeDict}[P]$;
\STATE ~~~ For each $P$ with the top-$k$ estimated score $\scoreapprox$,
\\     ~~~~~~ Compute the exact score $\score(P,\query)$ and \\ ~~~~~~ insert $P$ into the queue $Q$ (with size at most $k$);
\STATE Return the top-$k$ tree patterns in $Q$ and valid subtrees.
\end{algorithmic}
\caption{\algLinearEnumTopK$(\Lambda, \rho)$: partitioning by types and sampling roots to find the top-$k$ tree patterns}
\label{alg:linearenum:topk}
\end{algorithm}

\subsubsection{Speedup by Sampling}
\label{sec:algorithm:enum:sampling}

The two most costly steps in \algLinearEnum are in subroutine {\sc ExpandRoot}: i) the enumeration of tree patterns in the product of $\patns(\word_i, r)$'s (line~7); and ii) the enumeration of valid subtrees in the product of $\paths(\word_i, r, P_i)$'s (line~9). Too many valid subtrees could be generated and inserted into the dictionary ${\rm TreeDict}[]$ which is costly in both time and space. Now we introduce how to use sampling techniques to find the top-$k$ tree patterns more efficiently (but with probabilistic errors).

\stitle{Estimating scores using samples.} Instead of computing the valid subtrees for every root candidate (as {\sc ExpandRoot} in \calg\ref{alg:linearenum}), we do so only for a random subset of candidate roots -- each candidate root is selected with probability $\rho$. Equivalently, for each tree pattern $P$, only a random subset of valid subtrees in $\trees(P)$ are retrieved (kept in ${\rm TreeDict}[P]$), and we can use this random subset to estimate $\score(P,\query)$ as $\scoreapprox(P,\query)$. We then only maintain tree patterns with the top-$k$ estimated scores, without keeping the complete set of valid subtrees in $\trees(P)$ for each. Finally, we compute the exact scores and the complete sets of valid subtrees only for the estimated top-$k$, and re-rank them before outputting.

The detailed algorithm, called \algLinearEnumTopK, is described in \calg\ref{alg:linearenum:topk}. In addition to the input knowledge graph and keyword query, we have two more parameters $\Lambda$ and $\rho$. 
We first enumerate the type of roots in a tree pattern in line~2. For each type, similarly as \algLinearEnum, candidate roots of this are computed in line~3. We can compute the number of valid subtrees (possibly from different tree patterns) with these roots as $N_R$ in line~4, without really enumerating them. To this end, we only need to get the number of paths starting from each candidate root $r$ and ending at each keyword $\word_i$. Only when the number of valid subtrees is no less than $\Lambda$, we apply the root sampling technique in lines~7-8 with $rate = \rho$ (otherwise $rate = 1$): for each candidate root $r$, with probability $rate$, we compute the valid subtrees under it and insert them into the dictionary ${\rm TreeDict}[]$ (subroutine {\sc ExpandRoot} in \calg\ref{alg:linearenum} is re-used for this purpose). After all candidate roots of a type are considered, in lines~9-10, we can compute the estimated score as $\scoreapprox(P, \query)$ for each tree pattern $P$ in ${\rm TreeDict}$. Only for tree patterns with the top-$k$ estimated scores, we compute their valid subtrees with exact scores and insert them into a global queue $Q$ in line~11 to find the global top-$k$ tree patterns.

The running time of \algLinearEnumTopK can be controlled by parameters $\Lambda$ and $\rho$. Sampling threshold $\Lambda$ specifies for which types of roots, we sample the valid subtrees to estimate the pattern scores. By setting $\Lambda = +\infty$ and $\rho = 1$ (no sampling at all), we can get the exact top-$k$. When $\Lambda < +\infty$ and $\rho < 1$, the algorithm is speedup but there might be errors in the top-$k$ answers. In the experiments, we will show that even when $\rho = 0.1$ (\ie, use $10\%$ valid subtrees to estimate the pattern scores), we can get reasonably precise top-$k$ tree patterns while the algorithm is speedup roughly $10$ times. The theoretical analysis about the running time and precision of \algLinearEnumTopK are in the following two theorems.

\begin{theorem}\label{thm:linearenumtopk:time}
{\bf (Running Time)} For a keyword query $\{\word_1,$ $\ldots,$ $\word_m\}$ in a knowledge graph $\graph$, let $S_i$ be the size of the path index for word $\word_i$, let $N$ be the total number of valid subtrees, and let $|\types|$ be the total number of types. \algLinearEnumTopK needs time: \\$\bigoh{\min(\Lambda \cdot |\types|, N) \cdot d \cdot m + \rho \cdot N \cdot d \cdot m + \sum_{i=1}^m S_i + N \cdot \log k}$.

{\bf (Correctness)} When $\Lambda = +\infty$ and $\rho = 1$ (no sampling), the algorithm outputs the correct top-$k$ tree patterns.
\end{theorem}

We establish the {\em pairwise precision} of \algLinearEnumTopK: for two tree patterns $P_1$ and $P_2$ with exact scores $\score(P_1, \query)$ $>$ $\score(P_2, \query)$ in the general form of \eqref{equ:score:general}, how likely we would order them incorrectly, $\scoreapprox(P_1, \query) < \scoreapprox(P_2, \query)$, according to the estimated scores obtained from a random sample of valid subtrees (so that $P_1$ might be missed from the top-$k$ output by the algorithm).
%
%
%

\begin{theorem} \label{thm:cntpat:precision}
{\bf (Precision)} For a query $\query$ and tree patterns $P_1$ and $P_2$ with scores $\scoreexact_1 = \score(P_1, \query)$ and $\scoreexact_2 = \score(P_2, \query)$ s.t. $\scoreexact_1 > \scoreexact_2$, if \algLinearEnumTopK runs with $\Lambda = 0$ (always sampling) and sampling rate $\rho < 1$, then
%
%
$\scoreapprox(P_1,\query) < \scoreapprox(P_2,\query)$ ($P_1$ is incorrectly ranked lower than $P_2$ in estimation) with probability
\begin{equation} \label{equ:tailbound2}
\Pr[\hbox{error}] \leq \exp\left(-2\left(\frac{\scoreexact_1 - \scoreexact_2}{\scoreexact_1 + \scoreexact_2}\right)^2 \cdot \rho^2 \right).
\end{equation}
\end{theorem}
%

To prove the above theorem, we note that the score $\score(P_i, \query)$ can be decomposed among all candidate roots, \ie, rewritten as
\[
\scoreexact_i=\score(P_i, \query) = \!\!\!\!\!\!\!\! \sum_{T \in \trees(P_i)} \!\!\!\!\!\! \score(T, \query) = \sum_{r \in \entities} \sum_{T \in \trees_r(P_i)} \!\!\!\!\!\!  \score(T,q),
\]
where $\trees_r(P_i)$ is the set of valid subtrees with pattern $P_i$ and rooted at node $r$. Let $\scoreexact_i(r) = \sum_{T \in \trees_r(P_i)} \score(T,q)$ be the sum of relevance scores of all valid subtrees rooted at $r$ for pattern $P_i$, and thus $\scoreexact_i = \sum_r \scoreexact_i(r)$. In order to compare $\scoreexact_1$ and $\scoreexact_2$, we can compare $\sum_{r \in R^+} \scoreexact_1(r)$ and $\sum_{r \in R^+} \scoreexact_2(r)$ on a random subset $R^+$ of all candidate roots (sampled in line~8 of \algLinearEnumTopK with rate $\rho$). Using Hoeffding's inequality \cite{book:2009DubhashiP}, we can bound the probability that we make mistakes by a term that is exponentially small in the sampling rate $\rho$ and the difference between $\scoreexact_1$ and $\scoreexact_2$. Detailed proof can be found in the appendix.

The theorem has two direct implications which are consistent to our intuition:
%
%
i) the error probability decreases when the (relative) difference between $\score(P_1,\query)$ and $\score(P_2, \query)$ becomes larger; and ii) the error probability is smaller for higher sampling rate $\rho$ (exponentially in $\rho^2$).
They partly explain why the sampling technique works well in practice, as shown in \csec\ref{sec:exp:sampling}.

\stitle{How to set sampling threshold and sampling rate.} Intuitively, the sampling threshold $\Lambda$ determines {\em when to sample}, \ie, for each entity type, applying the sampling technique when the number of valid subtrees with roots of this type is no less than $\Lambda$; and the sampling rate $\rho$ determines the {\em sample size} for each root type. 

A global sampling threshold $\Lambda$ can be set regardless of the query and the number of valid subtrees w.r.t. it. The rationale is that, when the number of entity types is fixed, if the number of valid subtrees rooted in a type is less than $\Lambda$, sampling is not necessary (sampling rate set to 1 in line~5 of \calg\ref{alg:linearenum:topk}) because computing the exact scores is not expensive anyway. On the other hand, when the number of valid subtrees rooted in a type is at least $\Lambda$, we sample a fixed portion ($\rho$) of them to estimate the scores, and \cthm\ref{thm:cntpat:precision} provides a guarantee of precision w.r.t. $\rho$. So $\Lambda$ and $\rho$ can be set regardless of the queries, but they do rely on users' preference (trade-off between the response time of the system and the precision) for fixed scheme of the knowledge graph.
%

\remove{
[Working]

In this section, we describe four algorithms for the proposed problem. The first algorithm is a direct extension of the backward search algorithm. The second algorithm employs a simple optimization that partition the search space by the type of roots. The third algorithm tries to optimize the space complexity by materializing paths in the knowledge graph. The last algorithm tries to speedup the computation by sampling.

[summarize notations]

\subsection{Backward Search Based Algorithms}
[Create one node in the knowledge graph for each type.]

Our first algorithm is based on the backward search algorithm [citation] which finds all valid subtrees by exploring the knowledge graph starting from entities/types that contain keywords. We build an inverted index on the text description for entities. For a keyword $\word$, $inv_{\entities}(\word)$ is a list of entities such that $\word$ appears in the the text description of these entities. We also build inverted indices on the text description for types and attributes. Let $inv_{\types}(\word)$ and $inv_{\attributes}(\word)$ be the list of types and attributes that contain keyword $\word$ in their text description. The last preprocessing required by this algorithm is to construct a graph $\graph^{-1}$ that enables the backward traversal of the original knowledge graph. Each node in $\graph^{-1}$ is an entity, and there is a directed edge from $u$ to $v$ iff $v.A = u$ for some attribute $A \in \attributes(v)$.

The pseudocode for this backward search based algorithm is shown in \cfig\ref{alg:backwardSearch}. The algorithm generates all valid subtrees by enumerating the root of a subtree. It first builds a dictionary $idx$ by performing a depth-first search (DFS) starting from each entity/type that contains $\word_i$ in its text description. A key in $idx$ is a keyword $\word_i$ and a node $u$, and its value is a list of paths $u \rightsquigarrow v$ in $\graph$ such that $v.\textname$ contains $\word_i$. This list also includes all attributes of $u$ whose text description contains $\word_i$. Note that the minimality condition of a valid subtree, the DFS starting from $v$ ($v.\textname$ contains $\word_i$) doesn't search any further from a node $u$ even if the length of the path from $v$ to $u$ is less than $d$ when $u.\textname$ also contains $\word_i$. Given a node $u \in \entities$, if $idx[(\word_i, u)]$ is not empty for every keyword $\word_i$, i.e., there is a directed path from $u$ to an entity $v$ such that $\word_i$ appears in the the text description of a) $v$, b) the type of $v$, or c) the attribute of the edge pointing to $v$, then $u$ can be the root of a valid subtree. A such subtree $T$ can be constructed by picking one path from $idx[(\word_i, u)]$ for every $\word_i$. If $T$ is indeed a valid subtree, it is added into the dictionary $ans$ which groups valid subtrees by their patterns. Finally, the algorithm computes the relevance score of each pattern in $ans$, and returns the top-$k$ ranked patterns.

\begin{figure}[htbp]
\begin{algorithmic}[1]
\STATE {\bf for} $i := 1$ {\bf to} $m$
\STATE ~~ {\bf foreach} $v \in inv_{\types}(\word_i) \cup inv_{\entities}(\word_i)$
\STATE ~~ ~~ ~~ {\bf foreach} path $v \rightsquigarrow u$ of length at most $d$ obtained\\
 ~~ ~~ ~~ ~~ ~~ ~~ ~ from DFS on $\graph^{-1}$ starting from $v$
\STATE ~~ ~~ ~~ ~~ $idx[(\word_i, u)].add(u \rightsquigarrow v)$\\
 ~~ ~~ ~~ ~~ // $u \rightsquigarrow v$ is the corresponding path in $\graph$
\STATE ~~ {\bf foreach} $A \in inv_{\attributes}(\word_i)$
\STATE ~~ ~~ {\bf foreach} $u \in \entities$ that has attribute $A$
\STATE ~~ ~~ ~~ $idx[(\word_i, u)].add(u ~ A ~ u.A)$
\STATE {\bf foreach} $u \in \entities$
\STATE ~~ {\bf if} $idx[(\word_i, u)]$ is not empty $\forall i \in [1, m]$
\STATE ~~ ~~ {\bf foreach} subgraph $T$ constructed by picking one\\
~~ ~~ ~~ ~~ ~~ ~\; path from $idx[(\word_i, u)]$ $\forall i \in [1, m]$
\STATE ~~ ~~ ~~ {\bf if} $T$ is a valid subtree
\STATE ~~ ~~ ~~ ~~ $ans[\pattern(T)].add(T)$
\STATE {\bf return} patterns $P$ in $ans$ with the top-$k$ highest relevance scores
\end{algorithmic}
\caption{The backward search based algorithm.}
\label{alg:backwardSearch}
\end{figure}

\begin{example} \label{exa:backwardSearch}
Consider the keyword query in \cexp\ref{exa:validtree}. Table~\ref{tbl:backwardSearch} shows the keys and values in the dictionary $idx$ that are related to the valid subtrees $T_1$, $T_2$ and $T_3$. If the root is $v_1$, $T_1$ can be constructed from the values in row 1, 4, 7, 10 in the table. Similarly for $T_2$ and $T_3$.
\end{example}

\begin{table}[htbp]
\scriptsize
\center
\begin{tabular}{ll}
  \hline
  key & value \\ \hline
  ($\word_1$, $v_1$) & ($v_1$, Genre, $v_2$) \\
  ($\word_1$, $v_7$) & ($v_7$, Genre, $v_9$) \\
  ($\word_1$, $v_{12}$) & ($v_{12}$) \\
  ($\word_2$, $v_1$) & ($v_1$, $\type$, Software) \\
  ($\word_2$, $v_7$) & ($v_7$, $\type$, Software) \\
  ($\word_2$, $v_{12}$) & ($v_{12}$) \\
  ($\word_3$, $v_1)$ & ($v_1$, Developer, $v_3$, $\type$, Company) \\
  ($\word_3$, $v_7$) & ($v_7$, Developer, $v_8$, $\type$, Company) \\
  ($\word_3$, $v_{12}$) & ($v_{12}$, Publisher, $v_{13}$, $\type$, Company) \\
  ($\word_4$, $v_1$) & ($v_1$, Developer, $v_3$, Revenue, $v_4$) \\
  ($\word_4$, $v_7$) & ($v_7$, Developer, $v_8$, Revenue, $v_10$) \\
  ($\word_4$, $v_{12}$) & ($v_{12}$, Publisher, $v_{13}$, Revenue, $v_{14}$) \\
  $\vdots$ & $\vdots$ \\
  \hline
\end{tabular}
\caption{Keys and values in the dictionary $idx$ for the query in \cexp\ref{exa:validtree}.}\label{tbl:backwardSearch}
\end{table}


This algorithm keeps all valid subtrees in memory. Its space complexity is proportional to the number of valid subtrees. The algorithm has a very large memory footprint if there are millions of valid subtrees. Moreover, the ``group by" operation on millions of subtrees is very slow. An optimization is to partition the search space by the type of roots. The definition of ``table answer'' implies that, the roots of the subtrees with the same pattern are of the same type. Thus, we can replace the enumeration between line 8 - line 12 in \cfig\ref{alg:backwardSearch} with the operations in \cfig\ref{alg:typeOptimization}. For each type $C$, we create a dictionary that contains all valid subtrees whose roots are of type $C$. The patterns with the top-$k$ highest relevance scores are inserted into a priority queue of size $k$. The priority queue contains the global top ranked patterns when all types are enumerated. Now, the space complexity is proportional to the maximal number of valid subtrees whose roots are of a specific type. This number is usually smaller than the number of valid subtrees. However, this algorithm may still have a large memory footprint even for small values of $k$. In the following subsection, we will present an algorithm whose space complexity is proportional to $k$, i.e., an algorithm that has small memory footprint when the value of $k$ is small comparing to the number of valid subtrees.

\begin{figure}[htbp]
\begin{algorithmic}[1]
\STATE initialize a priority queue $Q$ of size $k$
\STATE {\bf foreach} $C \in \types$
\STATE ~~ {\bf foreach} entity $u$ of type $C$
\STATE ~~ ~~ {\bf if} $idx[(\word_i, u)]$ is not empty $\forall i \in [1, m]$
\STATE ~~ ~~ ~~ {\bf foreach} subgraph $T$ constructed by picking one\\
~~ ~~ ~~ ~~ ~~ ~~ ~~ path from $idx[(\word_i, u)]$ $\forall i \in [1, m]$
\STATE ~~ ~~ ~~ ~~ {\bf if} $T$ is a valid subtree
\STATE ~~ ~~ ~~ ~~ ~~ $ans[\pattern(T)].add(T)$
\STATE ~~ insert the patterns $P$ in $ans$ with the top-$k$ highest\\
~~ relevance scores to $Q$
\STATE {\bf return} patterns in $Q$
\end{algorithmic}
\caption{Partition the search space by the type of the roots.}
\label{alg:typeOptimization}
\end{figure}

\subsection{Path Materialization Algorithm}
In the previous algorithms, a table answer cannot be constructed until all valid subtrees have been generated. The algorithms have to keep all valid subtrees in memory during the computation. In this subsection, we consider an alternative approach called the Path Materialization algorithm that finds all valid subtrees with the same pattern at the same time. Thus, the algorithm can compute the relevance score of a pattern as soon as this pattern has been enumerated, and keep only top-$k$ ranked patterns in memory.

The algorithm materializes all paths up to length $d$ during the preprocessing. For a type $C$, it performs a DFS from each entity $u \in C$ up to depth $d$. These paths are grouped together by their patterns. Then an inverted index is built on the text description of the end nodes of these grouped paths. For a keyword $\word$, $inv_C(\word)$ is a list of grouped paths in the following form:
\[ \pattern(v_1^1 \rightsquigarrow v_l^1), (v_1^1, v_2^1, \cdots, v_l^1), (v_1^2, v_2^2, \cdots, v_l^2), \cdots, \]
where $\word$ appears in the text description of $v_l^1, v_l^2, \cdots$. Here, paths $v_1^1 \rightsquigarrow v_l^1$, $v_1^2 \rightsquigarrow v_l^2$, $\cdots$ have the same pattern. We put the shared pattern $P$ at the beginning of the list, and each remaining item in the list is a list of nodes in one path in this group. Let $list(inv_C(\word), P)$ be the sub-list starting from the second item in the list, i.e., a collection of nodes in this group. If the $\word$ appears in $C.\textname$ or $A.\textname$ for some $A \in \attributes(v)$ where $\type(v) \in C$, $inv_C(\word)$ includes the group of corresponding paths of length one. Note that the early termination optimization during the DFS mentioned in the backward search based algorithms can also be applied to this algorithm.

The pseudocode of the Path Materialization algorithm is shown in \cfig\ref{alg:pathMaterialization}. For a type $C$, it tries to find all possible combinations of path patterns such that there are some valid subtrees with these patterns. For a selected set of patterns $P_1, P_2, \cdots, P_m$, the list intersection operation finds a set of starting nodes $v_1^t$, such that there is a path starting from $v_1^t$ in every $list(inv_C(\word_i), P_i)$, i.e., $v_1^t$ can be the root of a valid subtree. If such nodes exist, then it constructs the corresponding subgraphs, keeps the valid subtrees, and computes the relevance score of this table answer. The algorithm finds all top ranked patterns when the enumeration for every type $C$ finishes.

\begin{figure}[htbp]
\begin{algorithmic}[1]
\STATE initialize a priority queue $Q$ of size $k$
\STATE {\bf foreach} $C \in \types$
\STATE ~~ enumerate all combinations of patterns (pick\\
~~ a pattern $P_i$ from $inv_C(\word_i)$ for each $i \in [1, m]$)
\STATE ~~ ~~ $R := \cap_{i = 1}^m list(inv_C(\word_i), P_i)$
\STATE ~~ ~~ {\bf if} $R \neq \emptyset$
\STATE ~~ ~~ ~~ construct subgraphs for nodes in $R$
\STATE ~~ ~~ ~~ construct a table answer from all valid subtrees
\STATE ~~ ~~ ~~ insert the pattern and the relevance score to $Q$
\STATE {\bf return} patterns in $Q$
\end{algorithmic}
\caption{The Path Materialization algorithm.}
\label{alg:pathMaterialization}
\end{figure}

\begin{example} \label{exa:pathMaterialization}
Consider the keyword query in \cexp\ref{exa:validtree}. Table~\ref{tbl:pathMaterialization} shows the keys and values in the inverted index constructed during the preprocessing that are related to the valid subtrees $T_1$, $T_2$ and $T_3$. The tree pattern $P_1$ in \cfig\ref{fig:treepattern1} can be constructed from the first four lines in Table~\ref{tbl:pathMaterialization}\subref{tbl:pathMaterialization-a}. $T_1$ and $T_2$ is generated during the construction for $P_1$. Similarly for $P_2$ and $T_3$.
\end{example}

\begin{table}[htbp]
\scriptsize
\center
\subfigure[Part of the inverted index for type Software.\label{tbl:pathMaterialization-a}]{
\begin{tabular}{ll}
  \hline
  key & value \\ \hline
  $\word_1$ & (($\type(v_1)$, Genre, $v_2$), ($v_1$, $v_2$), ($v_7$, $v_9$)) \\
  $\word_2$ & (($\type(v_1)$, $\type$, Software), ($v_1$, Software), ($v_7$, Software)) \\
  $\word_3$ & (($\type(v_1)$, Developer, $\type(v_3)$, $\type$, Company), ($v_1$, $v_3$, Company),\\
  & ($v_7$, $v_8$, Company)) \\
  $\word_4$ & (($\type(v_1)$, Developer, $\type(v_3)$, Revenue, $\type(v_4)$), ($v_1$, $v_3$, $v_4$),\\
  & ($v_7$, $v_8$, $v_{10}$)) \\
  $\vdots$ & $\vdots$ \\
\hline
\end{tabular}
}
\subfigure[Part of the inverted index for type Book.\label{tbl:pathMaterialization-b}]{
\begin{tabular}{ll}
  \hline
  key & value \\ \hline
  $\word_1$ & (($\type(v_{12})$), ($v_{12}$)) \\
  $\word_2$ & (($\type(v_{12})$), ($v_{12}$)) \\
  $\word_3$ & (($\type(v_{12})$, Publisher, $\type(v_{13})$, $\type$, Company),\\
  & ($v_{12}$, $v_{13}$, Company)) \\  
  $\word_4$ & (($\type(v_{12})$, Publisher, $\type(v_{13})$, Revenue, $\type(v_{14})$), ($v_{12}$, $v_{13}$, $v_{14}$)) \\
  $\vdots$ & $\vdots$ \\
  \hline
\end{tabular}
}
\caption{Keys and values in the inverted index in the Path Materialization algorithm for the query in \cexp\ref{exa:validtree}.}\label{tbl:pathMaterialization}
\end{table}

The space complexity of this algorithm is proportional to $k$ since it keeps at most $k$ table answers in memory. The price for this small memory footprint is two-fold. First, the size of the inverted index is much larger than that in the backward search based algorithms. Second, a combination of patterns may not correspond to any valid subtrees, while a combination of paths corresponds to a valid subtree w.h.p. in the backward search based algorithms. Moreover, the list intersection operation is time consuming. Thus, this algorithm may waste a significant amount of time on exploring combinations of patterns that don't lead to table answers. Our implementation adopts a DFS based strategy and tries to minimize the number of list intersection operations by pruning. A branch is pruned when $\cap_{i = 1}^k list(inv_C(\word_i), P_i)$ is empty for any $k \in [1, m]$.

\subsection{Speedup by Sampling}
The last algorithm we study explores sampling techniques to speedup the computation. In the backward search based algorithms, it is very time consuming to enumerate all possible subtrees when the number of subtrees constructed during the enumeration for a type $C$ is very big. Instead, the Sampling algorithm only enumerates a subset of all possible subgraphs.

The pseudocode for the Sampling algorithm is shown in \cfig\ref{alg:sample}. For a type $C$, the number of subtrees constructed during the enumeration is upper bound by $\sum_{\type(u) \in C} \prod_{i = 1}^m \|idx[(\word_i, u)]\|$. If this number if no less than a predefined threshold $t$ (i.e., there are a lot of subtrees whose roots are of type $C$), the algorithm enters sampling mode, otherwise it behaves exactly the same as the algorithm shown in \cfig\ref{alg:typeOptimization}. In the sampling mode, if a node $u$ is a candidate root of some valid subtrees, the enumeration for $u$ is skipped with probability $1 - p$ for a predefined sampling rate $p$. A dictionary $ans^{\prime}$ is built when the enumeration for $C$ finishes. Now, $ans^{\prime}[P]$ contains a subset of valid subtrees with pattern $P$. We estimate the relevance score of $P$ using subtrees in $ans^{\prime}[P]$, and finds the top-$k$ ranked patterns under this estimation. For each of the top ranked pattern $P$, we find the set of valid subtrees with pattern $P$ using the inverted index constructed in the Path Materialization algorithm, and then insert $P$ with its exact relevance score to $Q$. Finally, the priority queue $Q$ contains the global top ranked patterns when we repeat the previous procedure for every type.
        
\begin{figure}[htbp]
\begin{algorithmic}[1]
\STATE initialize a priority queue $Q$ of size $k$
\STATE {\bf foreach} $C \in \types$
\STATE ~~ {\bf if} $\sum_{\type(u) \in C} \prod_{i = 1}^m \|idx[(\word_i, u)]\| \geq t$
\STATE ~~ ~~ initialize a dictionary $ans^{\prime}$
\STATE ~~ ~~ {\bf foreach} entity $u$ of type $C$
\STATE ~~ ~~ ~~ {\bf if} $idx[(\word_i, u)]$ is not empty $\forall i \in [1, m]$
\STATE ~~ ~~ ~~ ~~ with probability $p$
\STATE ~~ ~~ ~~ ~~ ~~ {\bf foreach} valid subtree $T$ whose root is $u$
\STATE ~~ ~~ ~~ ~~ ~~ ~~ $ans^{\prime}[\pattern(T)].add(T)$
\STATE ~~ ~~ {\bf foreach} top-$k$ ranked pattern $P$ in $ans^{\prime}$
\STATE ~~ ~~ ~~ construct the exact table answer, insert $P$\\
~~ ~~ ~~ with the exact relevance score to $Q$
\STATE ~~ {\bf else} enumerate all subtrees as line 3 - line 8 in \cfig\ref{alg:typeOptimization}
\STATE {\bf return} patterns in $Q$
\end{algorithmic}
\caption{The Sampling algorithm ($t$ and $p$ are the parameters that control the sampling threshold and rate).}
\label{alg:sample}
\end{figure}        
}
%

\section{Experiments}
\label{sec:exp}

The following approaches for the $d$-height tree pattern problem are implemented in C\#. They are evaluated on a machine with 2.4 GHz Intel CPUs and 96 GB memory, under Windows Server.

\algBaseline: The baseline approach described in \csec\ref{sec:model:problem:adp}.

\algPE: Our first algorithm, the pattern enumeration-join approach \algPatternEnum described in \csec\ref{sec:algorithm:fast}.

\algLE: Our second algorithm, \algLinearEnumTopK, described in \csec\ref{sec:algorithm:enum}. Recall that, when the two parameters sampling threshold $\Lambda = +\infty$ and sampling rate $\rho = 1$, it gets exact top-$k$ answers; and otherwise, it gets approximate top-$k$.

\stitle{Datasets.} We compare the algorithms on two real-life datasets, \wiki \cite{url:wiki} and \imdb \cite{url:imdb}. The \wiki dataset contains 1.89 million entities. The type of each entity and its attributes are extracted from its infobox block on the top-right of its page. There are a total of 3,424 types. The corresponding knowledge graph contains 34.99 million edges.
%
%
The \imdb dataset contains 7 types
%
%
of 6.58 million entities, with 79.42 million directed edges in the knowledge graph.
%
%

\stitle{Queries.} We randomly selected 500 queries from Bing's log for experiments on \wiki. The numbers of keywords in the queries vary from 1 to 10, and for each we have 50 queries. For \imdb, we randomly constructed 500 queries from \imdb's vocabulary. Again, the numbers of keywords in the queries are from 1 to 10, and for each there are 50 queries. When we report the running time of an algorithm for a set of queries, we report the min / (geometric) average / max execution time in the form of error bars.
%

\stitle{Index size and height threshold $d$.} We build the path indexes described in \csec\ref{sec:algorithm:index} with different height thresholds $d = 2$, $3$, and $4$ for the \wiki dataset.
%
%
The time needed to construct them and their sizes are reported in \cfig\ref{tbl:index}. Both the time and the size increase exponentially in $d$ mainly because the number of possible tree patterns increases exponentially.
%
%
For \imdb, the knowledge graph contains only paths of length at most three, and the size of the indexes is 0.8 GB.
All the indexes are stored in memory.
\begin{figure}[t]
\center
\begin{tabular}{l|rrr}
  \hline
   & $d = 2$ & $d = 3$ & $d = 4$ \\
  \hline \hline
  Time (s) & 43 & 502 & 7,011 \\
  Size (MB) & 229 & 2,633 & 34,485 \\
  \hline
\end{tabular}
\vspace{-1em}
\caption{Index construction cost on Wiki for different $d$}
\vspace{-1.6em}
\label{tbl:index}
\end{figure}

\subsection{Performance of Exact Algorithms}
\label{sec:exp:exact}

We first compare different approaches when the exact top-$k$ tree patters are desired. No sampling is used in \algLE ($\Lambda = +\infty$ and $\rho = 1$). We use $k = 100$ by default and Exp-IV is about varying $k$.

\begin{figure}[t]
\center
\subfigure{
\includegraphics[width=0.48\textwidth]{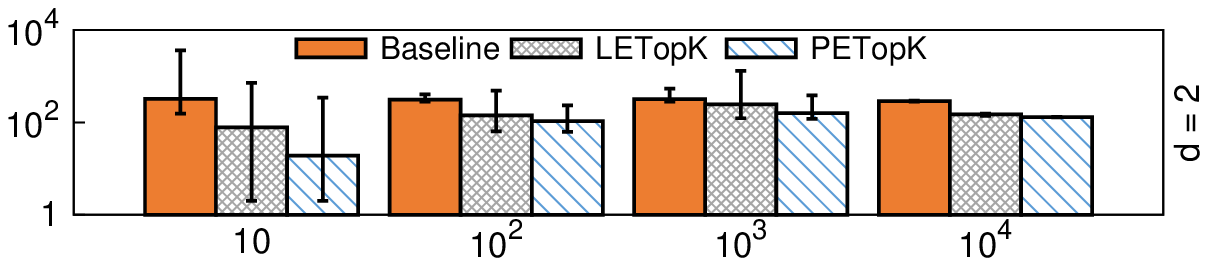}
\label{fig:wiki_time_vs_tree_d1}
}
\vspace{-2.4em}
\\
\subfigure{
\includegraphics[width=0.48\textwidth]{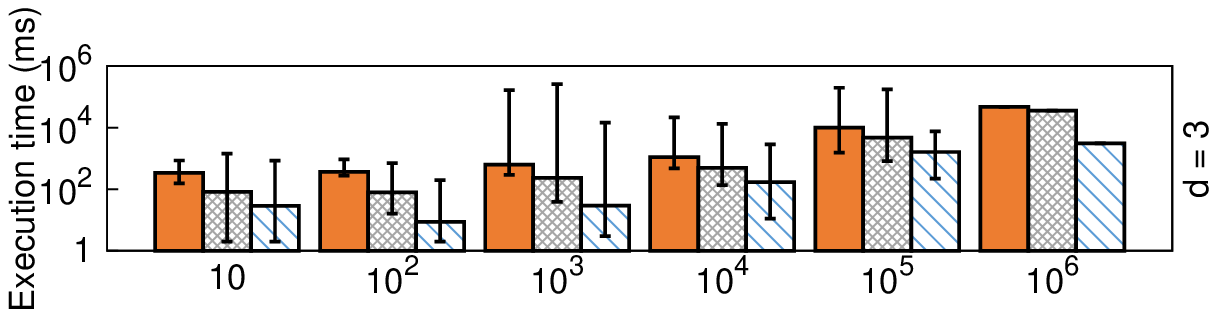}
\label{fig:wiki_time_vs_tree_d2}
\vspace{-2em}
}
\vspace{-2.4em}
\\
\subfigure{
\includegraphics[width=0.48\textwidth]{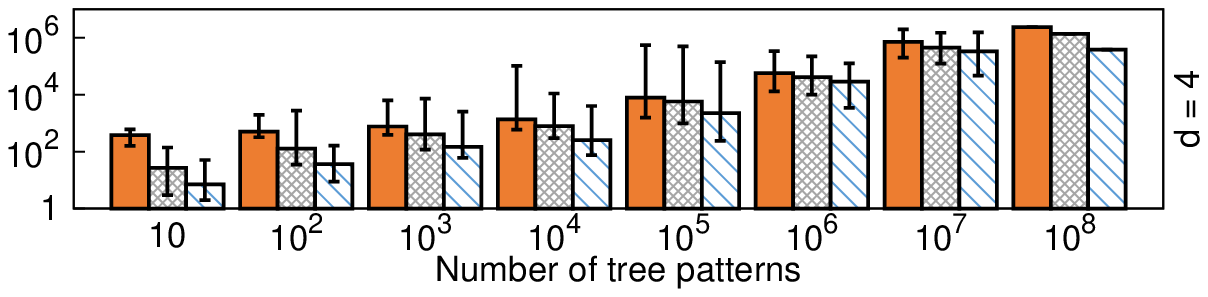}
\label{fig:wiki_time_vs_tree_d3}
}
\vspace{-2.4em}
\caption{Execution time and number of tree patterns (with height at most $d$) for different height threshold $d$ on \wiki}
\vspace{-1.2em}
\label{fig:wiki_time_d}
\end{figure}

\begin{figure}[t]
\center
\includegraphics[width=0.48\textwidth]{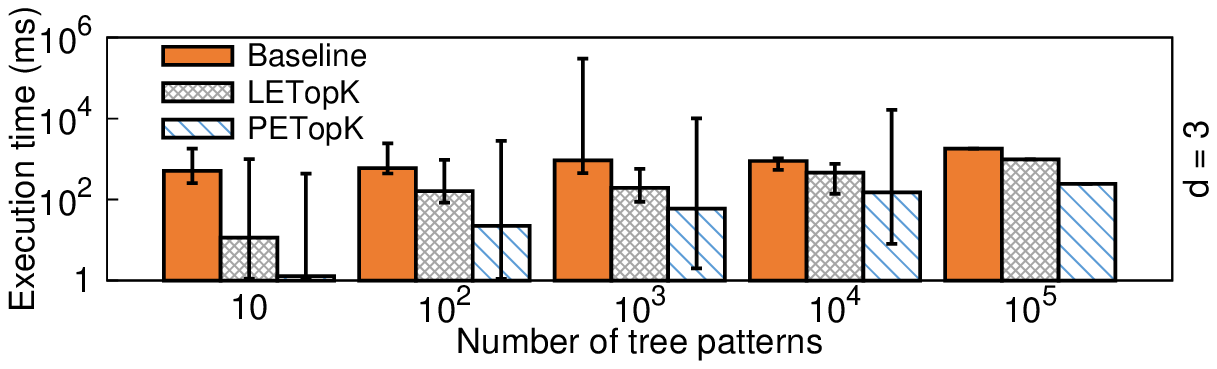}
\vspace{-2.4em}
\caption{Execution time and number of tree patterns with height at most $d$ on \imdb}
\vspace{-1.6em}
\label{fig:imdb_time_d}
\end{figure}

\stitle{Exp-I: Varying height threshold $d$ and number of tree patterns.} We first vary the height threshold $d$ for the \wiki dataset. When $d$ increases, the number of paths with length at most $d$ increases significantly,
%
%
and as a result, for a fixed query, the number of valid subtrees and tree patterns also increase significantly. From \cfig\ref{fig:wiki_time_d}, we can see that the number of tree patterns increases from $[10, 10^4]$ to $[10, 10^8]$ for $d = 2, 3, 4$.
For each $d$, we study how the number of tree patterns affects the execution time of query processing. The 500 queries on \wiki are partitioned into different groups based on the total number of possible tree patterns that can be found for each query, \eg, group $10^2$ contains all queries with $10$ -- $99$ tree patterns. The results are reported in \cfig\ref{fig:wiki_time_d}.

It can be seen that larger $d$ greatly affects the performance of our algorithms, with a larger number of possible tree patterns as the major reason. Overall, \algLE is faster than \algBaseline, and \algPE is the fastest among the three algorithms.
%
%
We want to emphasize that the advantage of \algLE in practice mainly relies on the sampling technique. But sampling is disabled for now to compare exact top-$k$ algorithms, and will be discussed in \csec\ref{sec:exp:sampling}.

In terms of the answer quality, on one hand, $d$ should be large enough to ensure that we explore enough number of interpretations for the query; and on the other hand, if $d$ is too large, some large tree patterns that correspond to loose relationship among keywords may appear among the top answers, which actually deteriorate the answer quality. Similar finding was also made in \cite{sigmod:LiOFWZ08} for ranking individual subtrees. In our case, when $d=3$, the best interpretations (tree patterns) of the queries on \wiki can be found at an average ranking of $2.797$. We will miss some of them for $d=2$. But for $d=4$, the (same) best interpretations have an average ranking of $12.514$.
%
%
So we use $d=3$ in the rest experiments for \wiki.

In \imdb, the max length of directed paths is three, so $d=3$ suffices (since tree patterns here have heights at most 3). The results are reported in \cfig\ref{fig:imdb_time_d} for $d=3$. The set of answers and execution time will be exactly the same for $d > 3$. Similar to the results in \wiki, while the number of possible tree patterns affects execution time, \algPE is the fastest one on average.

\stitle{Exp-II: Varying number of valid subtrees.}
Besides the number of tree patterns, another important parameter about a keyword query is how many valid subtrees in total can be found in the knowledge graph. This parameter may affect the performance of algorithms a lot. For example, \cthm\ref{thm:linearenum:correct} indicates that the running time of \algLE is linear in this number. 
%
%
So we partition queries into different groups based on how many valid subtrees a query has in total (\eg, group $10^3$ contains all queries with 100 -- 999 valid subtrees).
%
\cfig\ref{fig:time_vs_tree} reports the execution time when varying the number of valid subtrees on both \wiki and \imdb.
%
%

\begin{figure}[t]
\center
\subfigure[Varying number of valid subtrees on \wiki dataset]{
\includegraphics[width=0.48\textwidth]{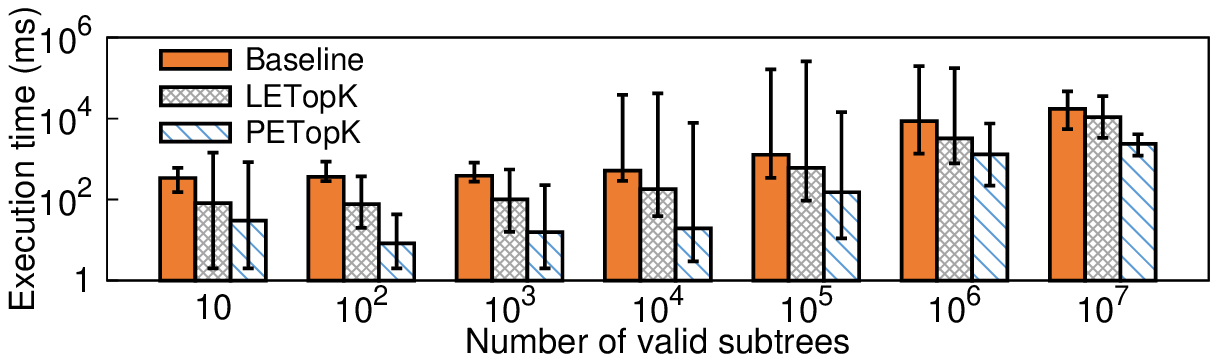}
\label{fig:wiki_time_vs_tree}
}
\subfigure[Varying number of valid subtrees on \imdb dataset]{
\includegraphics[width=0.48\textwidth]{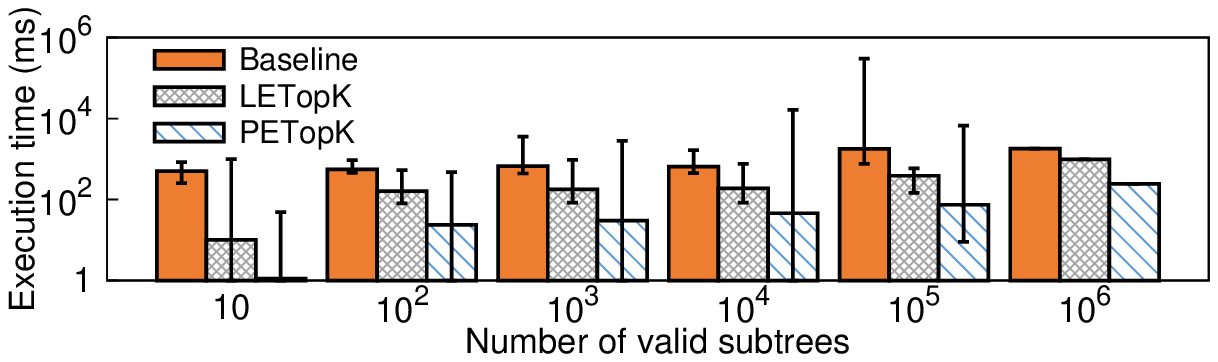}
\label{fig:imdb_time_vs_tree}
}
\vspace{-2em}
\caption{Execution time for different queries}
\vspace{-1em}
\label{fig:time_vs_tree}
\end{figure}


Again, \algLE is faster than \algBaseline, and \algPE is the fastest among the three algorithms. The execution time of \algBaseline and \algLE is bound by the time on building the dictionary ${\rm TreeDict}$. \algLE is faster than \algBaseline as a result of the ``partitioning by types'' technique in \csec\ref{sec:algorithm:enum:topk}. \algPE is usually the fastest since the pattern-first path index it uses allows it to avoid the time consuming dictionary building and online aggregation.
%


\stitle{Exp-III: Varying size of knowledge graph.}
We study the scalability of different algorithms on the \wiki dataset by varying the number entities and types in the knowledge graph. We randomly select a subset of entities from the \wiki dataset, and construct the induced subgraph of the original knowledge graph w.r.t. the selected subset of entities. The execution time of each algorithm on the induced knowledge graphs for different numbers (10\%-100\%) of entities is shown in \cfig\ref{fig:wiki_scale_entity}. The execution time of each algorithm increases (almost) linearly as the number of entities increases from 10\% to 100\% of the entities in the \wiki dataset.

Similar results are found for varying numbers of entity types in the knowledge graph. Details are omitted for the space limit.
%

\begin{figure}[t]
\center
\includegraphics[width=0.48\textwidth]{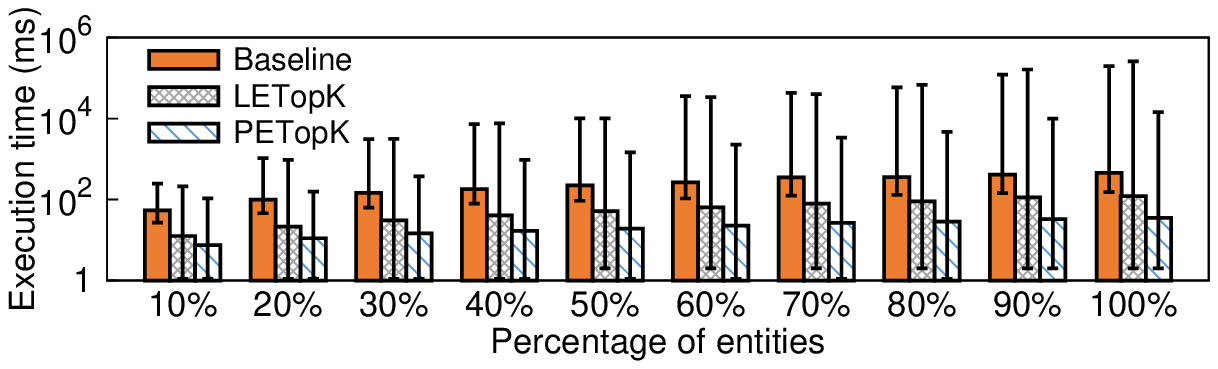}
\vspace{-2em}
\caption{Execution time on \wiki datasets of different sizes}
\label{fig:wiki_scale_entity}
\vspace{-1em}
\end{figure}

\remove{
\begin{table}[ht]
\center
\begin{tabular}{l|rrr}
  \hline
  Algorithm & $d = 1$ & $d = 2$ & $d = 3$ \\
  \hline \hline
  \algLinearEnum & 437.99 & 1,033.75 & 17,008.27 \\
  \algLinearEnumTopK & 157.32 & 415.98 & 7,027.25 \\
  \algPatternEnum & 21.95 & 158.20 & 3,981.29 \\
  \hline
\end{tabular}
\vspace{-1em}
\caption{(Geometric) Average execution time (s).}
\label{tbl:wiki_time_d}
\end{table}
}

\stitle{Exp-IV: Varying parameter $k$.}
The value of $k$ has very little impact on the execution time of our algorithms. For each tree pattern, it takes $\bigoh{\log k}$ operations to insert it to the priority queue of size $k$, while the number of operations required to find it is independent of $k$ (which is usually much larger than $\bigoh{\log k}$). Thus, the execution time is dominated by the aggregation/enumeration of valid subtrees, and is almost not affected by the value of $k$.

\remove{
\begin{figure*}
\begin{tabular}{|l|l|l|}
\hline
{\bf company} & genre & {\bf founder}\\ \hline
MySQL AB & {\bf Database} & Michael Widenius, David Axmark and Allan Larsson \\
Sleepycat Software & {\bf Database} & Margo Seltzer and Keith Bostic \\ 
... & ... & ... \\ \hline
\end{tabular}
\\
\\
\\
\\
\begin{tabular}{|l|l|l|l|l|}
\hline
dot-com {\bf company} & programming language & developer & product & {\bf founder}\\ \hline
Amazon.com & Java & Oracle Corporation & Oracle {\bf Database} & Jeff Bezos \\
EBay & Java & Oracle Corporation & Oracle {\bf Database} & Pierre Omidyar \\
... & ... & ... & ... & ... \\ \hline
\end{tabular}
\end{figure*}
}

\remove{
\begin{figure*}[ht]
\center
\begin{minipage}{0.576\textwidth}
\subfigure[Query 1.]{
\includegraphics[width=0.34\textwidth]{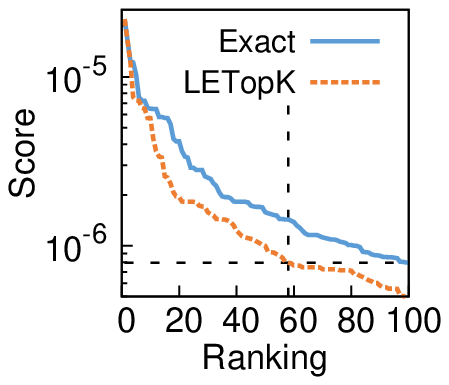}
\label{fig:score_q1}
}
\hspace{-2.2em}
\subfigure[Query 2.]{
\includegraphics[width=0.34\textwidth]{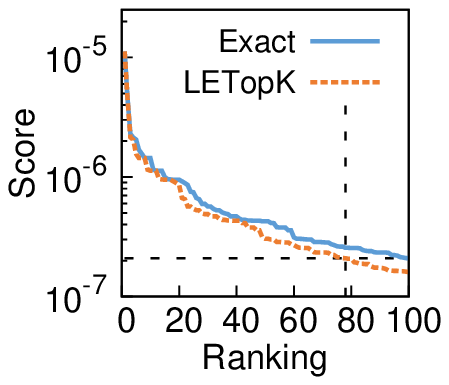}
\label{fig:score_q2}
}
\hspace{-2.2em}
\subfigure[Query 3.]{
\includegraphics[width=0.34\textwidth]{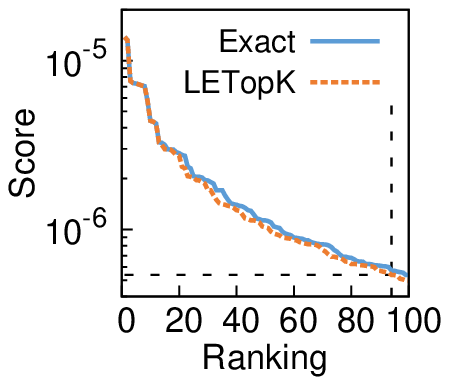}
\label{fig:score_q3}
}
\vspace{-1em}
\caption{Scores of exact top-$k$ tree patterns and Scores of top-$k$ found by \algLE ($\Lambda=10^5$, $\rho = 0.01$, $k = 100$).}
\label{fig:score_q123}
\end{minipage}
\hspace{1em}
\addtocounter{figure}{1}
\begin{minipage}{0.384\textwidth}
\subfigure[Execution time]{
\includegraphics[width=0.49\textwidth]{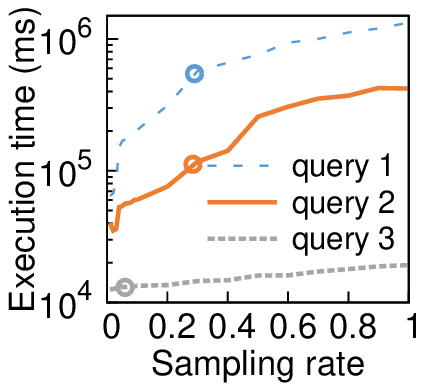}
\label{fig:wiki_sample_time_vs_rate}
}
\hspace{-1.4em}
\subfigure[Precision]{
\includegraphics[width=0.49\textwidth]{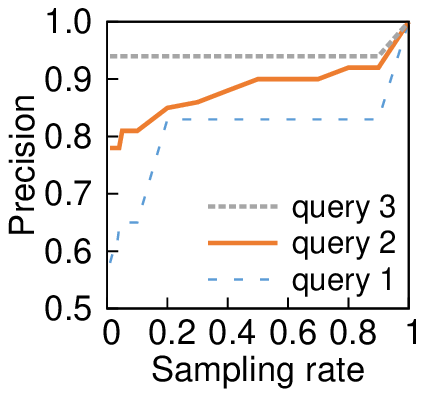}
\label{fig:wiki_sample_precision_vs_rate}
}
\vspace{-1.4em}
\caption{Performance of \algLE with different sampling rate ($\Lambda = 10^5$, $k = 100$). The circles on curves in \cfig\ref{fig:wiki_sample_time_vs_rate} mark the execution time of \algPE.}
\label{fig:wiki_sample_rate}
\end{minipage}
\end{figure*}
}

\subsection{Performance of Sampling Algorithm}
\label{sec:exp:sampling}

Now we study the performance of the sampling technique used in \algLE. Execution time and precision are the two measures that we are interested in. The {\em precision} here is defined as {\em the ratio between the number of truely top-$k$ answers found by \algLE (with sampling) and $k$}.
%
%
We focus on \wiki, since the number of valid subtrees is usually much smaller on \imdb (so sampling is not useful there). We selected three queries with different numbers of valid subtrees. The numbers of valid subtrees / tree patterns for them are (2,479,899 / 314,614), (819,739 / 61,967) and (540,849 / 32,300).

\begin{figure}[t]
\center
\subfigure{
\includegraphics[width=0.48\textwidth]{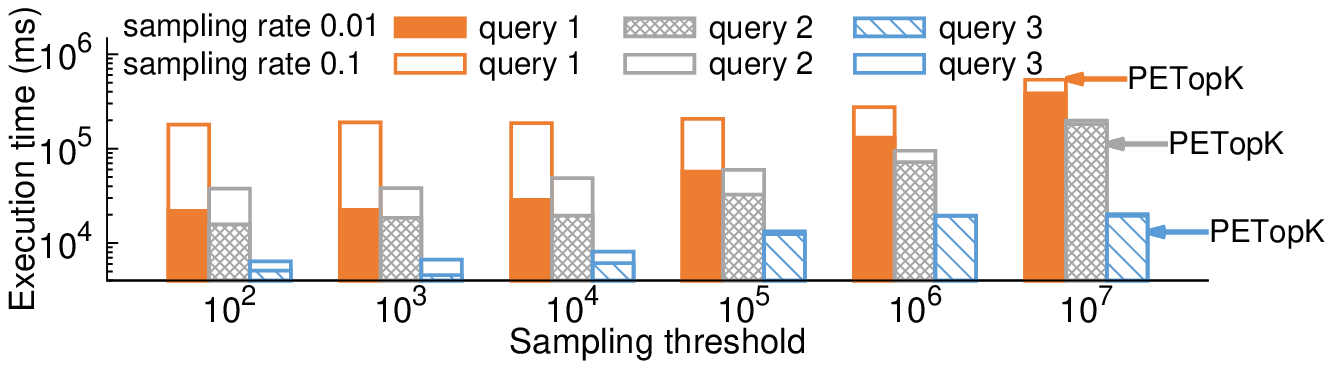}
}
\\\vspace{-0.6cm}
\subfigure{
\includegraphics[width=0.48\textwidth]{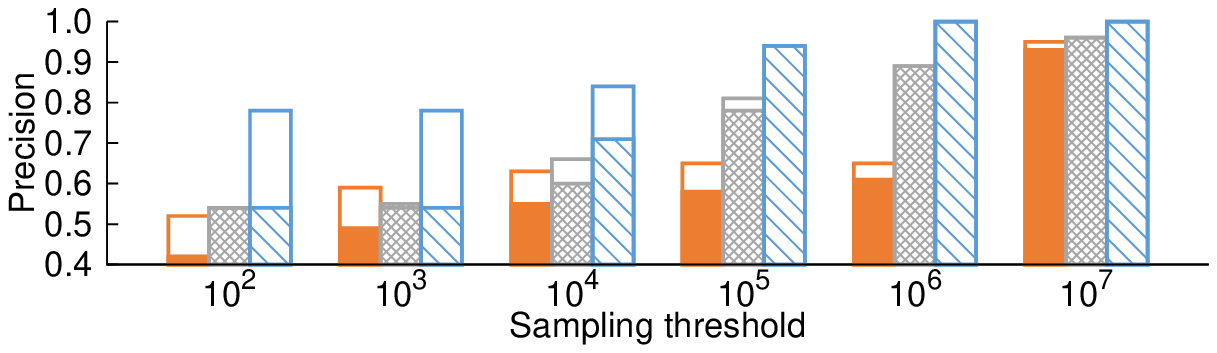}
}
\vspace{-2em}
\caption{Performance of \algLE with different sampling threshold $\Lambda$ (for $k = 100$). The execution time of \algPE is marked on the side of histograms.}
\vspace{-1.6em}
\label{fig:wiki_sample_threshold}
\end{figure}

\begin{figure}[t]
\centering
\subfigure[Execution time]{
\includegraphics[width=0.2\textwidth]{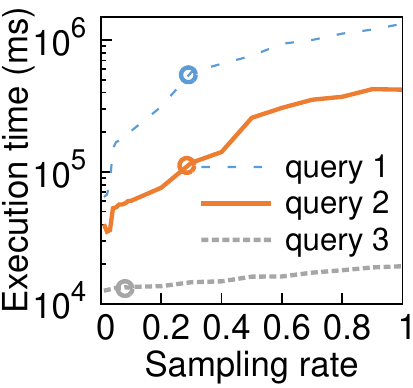}
\label{fig:wiki_sample_time_vs_rate}
}
\subfigure[Precision]{
\includegraphics[width=0.2\textwidth]{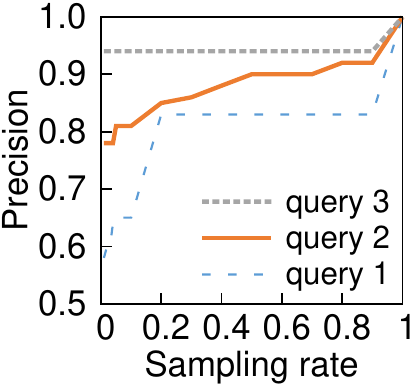}
\label{fig:wiki_sample_precision_vs_rate}
}
\vspace{-1.4em}
\caption{Performance of \algLE with different sampling rate $\rho$ (for $\Lambda = 10^5$, $k = 100$). The circles on curves in \cfig\ref{fig:wiki_sample_time_vs_rate} mark the execution time of \algPE.}
\vspace{-1.6em}
\label{fig:wiki_sample_rate}
\end{figure}

\remove{
\stitle{Exp-V: Overall performance.}
We run \algLE with $\Lambda = 10^5$ and $\rho = 0.01$ on the selected queries to find the top-$k$ tree patterns. The scores of the exact top-$k$ tree patterns and the scores of the top-$k$ found by \algLE are both shown in \cfig\ref{fig:score_q123}. We can see that the two curves are close to each other especially at the beginning part (for top answers). The scores follow a distribution with a ``long tail''. In each subfigure of \cfig\ref{fig:score_q123}, the vertical dotted line marks the number of correct answers found by \algLE. If the number is $c$, then the top ranked $c$ tree patterns are all correct, and the remaining $k - c$ tree patterns are incorrect. \algLE fails to find a correct answer (tree pattern) if 1) none of the roots in the tree pattern are sampled or 2) the estimated score of the tree pattern is not top-$k$ ranked. Ideally, both situations can be avoided if the sampling threshold (or the sampling rate) is high enough. But the execution time also increases as the sampling threshold (or sampling rate) increases. Thus, a sampling threshold (or sampling rate) with high precision and short execution time is favored.
}

\stitle{Exp-V: Varying sampling threshold $\Lambda$.}
Recall that the sampling threshold $\Lambda$ determines {\em when to sample}: for each entity type, applying the sampling technique when the number of valid subtrees with roots of this type is no less than $\Lambda$; and the sampling rate $\rho$ determines the {\em sample size} for each root type.
The performance of \algLE for different sampling threshold is reported in \cfig\ref{fig:wiki_sample_threshold} for $\rho = 0.01$ and $0.1$. Overall, both the execution time and the precision increase when the sampling threshold increases. We mark the execution time of \algPE in \cfig\ref{fig:wiki_sample_threshold}. \algLE is slower than \algPE for a very large sampling threshold (e.g., $10^7$) but becomes faster when the threshold is less than or equal to $10^5$. In the next experiment, we will fix $\Lambda = 10^5$ (as a balance between efficiency and precision), and vary the sampling rate.

\stitle{Exp-VI: Varying sampling rate $\rho$.}
The performance of \algLE for different sampling rate is in \cfig\ref{fig:wiki_sample_rate}. The circle on the execution time curve for each query is the execution time of \algPE.
%

For queries with larger numbers of valid subtrees (query 1 and query 2), \algLE becomes much (5x-20x) faster than \algPE when a smaller sampling rate is used (e.g., 0.2 for query 1 and query 2), while preserving reasonably high precision (above 80\%).
%

%
For the query with a smaller number of valid subtrees (query 3), the performance of \algPE is on a par with \algLE. The reason is that, for \algLE, the sampling threshold $\Lambda = 10^5$ is large in comparison to the number of valid subtrees (540,849 for query 3) -- so only for a few entity types where the numbers of valid subtrees rooted are larger than $10^5$, the sampling technique is enabled. As a result, only when the sampling rate used in \algLE is small enough ($\leq 0.05$), \algLE is faster than \algPE, but the precision of $\algLE$ is still consistently stable at round 0.95 (because sampling is enabled only for a small number of root types).

\remove{
\addtocounter{figure}{-2}
\begin{figure}[t]
\center
\subfigure[Execution time]{
\includegraphics[width=0.48\textwidth]{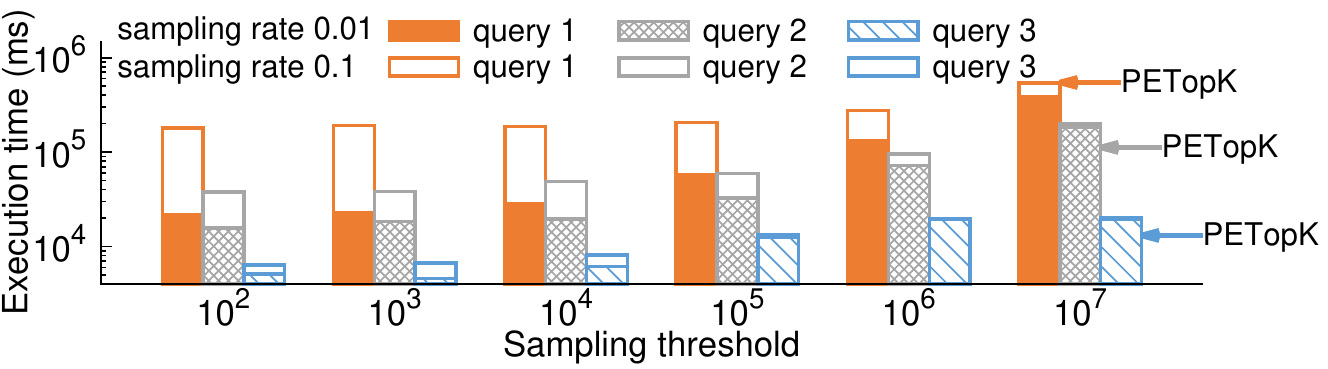}
\label{fig:wiki_sample_time_vs_threshold}
}
\\
\subfigure[Precision]{
\includegraphics[width=0.48\textwidth]{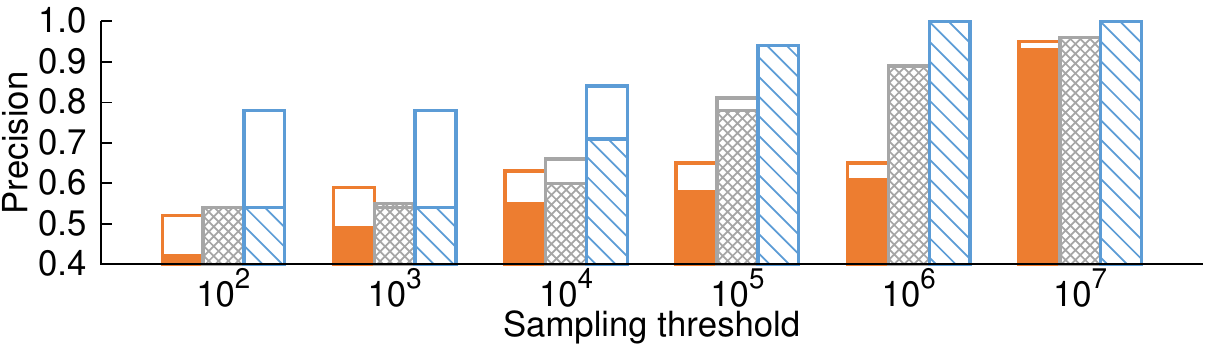}
\label{fig:wiki_sample_precision_vs_threshold}
}
\vspace{-2em}
\caption{Performance of \algLE with different sampling threshold ($k = 100$).}
\vspace{-2em}
\label{fig:wiki_sample_threshold}
\end{figure}
\addtocounter{figure}{2}
}

\subsection{Individual Trees v.s. Tree Patterns}

Recall the major motivation of this paper is to search tree patterns when the users want to find table answers (each represented as a set of subtrees with the same tree pattern) using keywords. We are not excluding individual best valid subtrees. But we aim to provide an additional module for the search engine to produce and rank highly relevant tree patterns (table answers). This new module could co-exist with the individual-page ranking module or individual-tree ranking module. Which module we want the search engine to direct users to automatically according to the query intention analysis and how to mix individual valid subtrees with tree patterns to provide a universal ranking are both open problems. It will be an interesting future work to address them using extensive user study.

\begin{figure}[t]
\centering
\subfigure{
\includegraphics[width=0.2\textwidth]{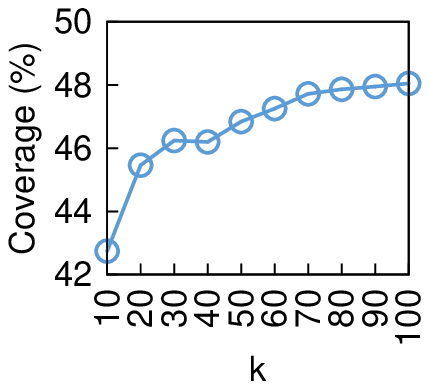}
}
\subfigure{
\includegraphics[width=0.2\textwidth]{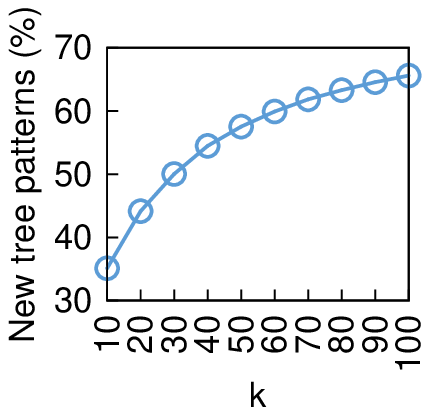}
}
\vspace{-1.2em}
\caption{Average coverage of individual relevant trees in top-$k$ tree patterns, and new tree patterns found in top-$k$}
\vspace{-1.6em}
\label{fig:individual_coverage}
\end{figure}

%

We compute a separate list of individual top-$k$ valid subtrees, based on their tree scores in \cequ\eqref{equ:score:tree}.
For the 500 keyword queries on \wiki, we calculate the average coverage of the individual top-$k$ subtrees in top-$k$ tree patterns (each as one row in some aggregated table), and the average percentage of top-$k$ patterns that cannot be found in the individual top-$k$ subtrees. The results are reported in \cfig\ref{fig:individual_coverage}, for $k$ varying from 10 to 100.
Because of their ``singular'' patterns (\ie, only a small number of valid subtrees have the same pattern), around half of the individual top-$k$ subtrees are lost in the top-$k$ tree patterns. At the same time, up to 70\% of the top-$k$ tree patterns are new to the individual top-$k$ subtrees.
%

\stitle{Case study.} We consider the query ``XBox Game'' in the \wiki dataset to compare the individual top valid subtrees and the top tree patterns in \cfigs\ref{fig:topk_ind_1}-\ref{fig:topk_pat_1}. Both individual subtrees and tree patterns are shown as tables with column names as edge(attribute)/node types and row cells as entities.
The top-1 individual valid subtree for ``XBox Game'' finds the entity ``XBox'', because of its relatively high PageRank score, with one additional edge/attribute containing the keyword ``game''. The top-2 finds a bigger subtree with ``DVD'' as the root and two branches ``DVD-usage-XBox'' and ``DVD-owners-Sony-products-video game'', and it ranks high mainly because of the high PageRank score of ``DVD''. The top-3 finds a singular entity with ``XBox'' appearing in the entity name and ``Game'' in the entity type.
Of course, when the user's intention is to find ``a list of XBox games'' by issuing this query, the tree pattern/table answer shown in \cfig\ref{fig:topk_pat_1} is better; and when the intention is to find ``popular XBox game'', the top-1 individual valid subtree in \cfig\ref{fig:topk_ind_1} is also a good candidate. Top-2 and top-3 valid subtrees are the cases when a top individual subtree is lost in our top-$k$ tree pattern answers because of the singularity of its pattern (no other valid subtree has the same pattern).

\begin{figure}[t]
\scriptsize
Top-1
\begin{tabular}{|l|l|}
\hline
information appliance & top {\bf game}\\ \hline\hline
{\bf Xbox} & Halo 2 \\ \hline
\end{tabular}
\\Top-2
\begin{tabular}{|l|l|l|l|}
\hline
storage medium & usage & owners/creators & products\\ \hline\hline
DVD & {\bf Xbox} & Sony & video {\bf game}\\ \hline
\end{tabular}
\\Top-3
\begin{tabular}{|l|}
\hline
video {\bf game} online service\\ \hline\hline
{\bf Xbox} Live Arcade \\ \hline
\end{tabular}
\vspace{-1em}
\caption{Top individual valid subtrees for ``XBox Game''}
\vspace{-1.4em}
\label{fig:topk_ind_1}
\end{figure}
\begin{figure}[t]
\scriptsize
Top-1
\begin{tabular}{|l|l|}
\hline
video {\bf game} & platform\\ \hline\hline
Halo 2 & {\bf Xbox} \\
GTA: San Andreas & {\bf Xbox} \\
Painkiller & {\bf Xbox} \\
... & ... \\ \hline
\end{tabular}
\vspace{-1em}
\caption{Top-$1$ tree pattern for ``XBox Game''}
\vspace{-2.2em}
\label{fig:topk_pat_1}
\end{figure}

\remove{
The query ``Country Capital'' is the case when the top-$k$ individual subtrees are not lost but are combined and provided to users as one answer by our tree-pattern search algorithms. The top-3 individual valid subtrees have the same pattern and can be combined as a table which can be consumed by users more easily (in \cfigs\ref{fig:topk_ind_2}-\ref{fig:topk_pat_2}).

\begin{figure}[t]
\scriptsize
Top-1
\begin{tabular}{|l||l|}
\hline
{\bf country} & {\bf capital}\\ \hline
United States & Washington, D.C. \\ \hline
\end{tabular}
\\Top-2
\begin{tabular}{|l||l|}
\hline
{\bf country} & {\bf capital}\\ \hline
England & London \\ \hline
\end{tabular}
\\Top-3
\begin{tabular}{|l||l|}
\hline
{\bf country} & {\bf capital}\\ \hline
Canada & Ottawa \\ \hline
\end{tabular}
\caption{Subtrees for ``Country Capital''}
\label{fig:topk_ind_2}
\end{figure}
\begin{figure}[t]
\scriptsize
Top-1
\begin{tabular}{|l||l|}
\hline
{\bf country} & {\bf capital}\\ \hline
United States & Washington, D.C. \\
England & London \\
Canada & Ottawa \\
... & ... \\ \hline
\end{tabular}
\caption{Top-$1$ tree pattern for ``Country Capital''}
\label{fig:topk_pat_2}
\end{figure}
}

\remove{
\subsection*{Summary} The two approaches \algPE and \algLE we proposed outperform the baseline algorithm in all the different settings. Both \algPE and \algLE scale well for knowledge graphs of different sizes, and for queries with different numbers of matching tree patterns. Although with worse theoretical guarantee, \algPE performs even better than \algLE in most cases. However, \algLE can be speed up using the sampling technique introduced in \csec\ref{sec:algorithm:enum:sampling}, which makes it more suitable for larger queries (5x-20x faster than \algPE while preserving precision over 80\%). We also show that a significant portion of individually highly relevant subtrees will be covered in our highly relevant tree patterns, based on the aggregate results and case studies.
}

%

\section{Related Work}
\label{sec:related}

\stitle{Searching and ranking tables.}
As search engines are able to keep more and more tables from the Web, there have been efforts to utilize these tables. On one hand, Web tables can be leveraged and returned directly as answers in response to keyword queries \cite{pvldb:LimayeSC10, pvldb:VenetisHMPSWMW11, pvldb:PimplikarS12, sigmod:YakoutGCC12}. On the other hand, Web tables can be used to understand keyword queries better through mapping query words to attributes of tables \cite{sigmod:SarkasPT10} and to provide direct answers to fact lookup queries \cite{www:YinTL11}.
Different from the above works, in this paper, we focus on the scenarios when relevant and complete tables are not available for user-given keyword queries, and our goal is to compose tables online as answers to those queries.

\stitle{Searching subtrees/subgraphs in RDB.}
Previous studies on keyword search in RDB extend ranking documents/webpages into ranking substructures of joining tuples which together contain all keywords in a query. They model an RDB as a graph, where tuples/tables are nodes and foreign-key links are edges. Each answer to a keyword query in such a graph could be either a subtree (\cite{icde:AgrawalCD02}, \cite{vldb:HristidisP02}, etc.) or a subgraph (\cite{sigmod:QinYC09}, \cite{sigmod:LiOFWZ08}, etc.) with all the keywords contained. There are two lines of works with the same goal of finding and ranking these answers. The first line materializes the RDB graph and proposes indexes and/or algorithms to enumerate top-$k$ subtrees or subgraphs \cite{icde:BhalotiaHNCS02, vldb:KacholiaPCSDK05, pods:KimelfeldS06, icde:DingYWQZL07, sigmod:HeWYY07, sigmod:GolenbergKS08, sigmod:LiOFWZ08}, etc. The second line first enumerates possible join trees/plans (candidate networks) based on the database schema and then evaluates them using SQL to obtain the answers \cite{icde:AgrawalCD02, vldb:HristidisP02, vldb:HristidisGP03, sigmod:LiuYMC06, sigmod:LuoLWZ07, sigmod:QinYC09, tkde:LuoWLZWL11}, etc. Yu et al. \cite{survey:2010Yu} provide a comprehensive survey on these two lines.

Our enumeration-aggregation baseline borrows ideas from the first line of previous works. It essentially first enumerates valid subtrees in our knowledge graph and groups them by their tree patterns. But this method is deficient because the bottleneck now is the grouping step instead of the enumeration step.
The second line of works (candidate network enumeration-evaluation) are not applicable in our problem because the schema of a knowledge base is usually much larger than the schema of an RDB, and thus the first step, {\em candidate network enumeration}, becomes the bottleneck. \cite{pods:KimelfeldS11} analyzes the complexity of this subproblem and proposes a novel parameterized algorithm which is interesting in theory.

\stitle{Keyword search in XML data.} Another important line of works are to search LCAs (lowest common ancestors) in XML trees using keywords, \cite{vldb:LiYJ04, sigmod:XuP05, www:SunCG07, pvldb:LiuC08}, etc. The general goal is to find lowest common ancestors of groups of nodes containing the keywords in the query. These LCAs, together with keyword-node matches sometimes, are returned as answers to the keyword query. Various strategies to identify relevant matches by imposing constraints on answers are developed, such as {\em meaningful LCA} \cite{vldb:LiYJ04}, {\em smallest LCA} \cite{sigmod:XuP05, www:SunCG07}, and {\em MaxMatch} \cite{pvldb:LiuC08}. LCA-based approaches are not applicable in our problem for two reasons: i) our goal is to find and rank tree patterns, each of which aggregates a group of subtrees according to the node/edge types on the paths from the root to each leaf containing a keyword, instead of individual roots/matches as in LCA approaches -- if we enumerate all LCA matches and group them by patterns, it would be equivalent to our baseline; and ii) LCA is not well defined in our knowledge graph with cycles.

In addition, XSeek \cite{sigmod:LiuC07} tries to infer users' intention by categorizing keywords in the query into {\em predicates} and {\em return nodes}. And \cite{tods:LiuC10} defines an equivalence relationship among query results on XML based on the classification of predicates and return nodes of keywords. \cite{www:LiuC11} provides a comprehensive survey on this line.

\stitle{Keyword search in RDF graphs.}
Our knowledge graph can be considered as an RDF graph. Previous works on keyword search in RDF graphs extend the two lines of works on keyword search in RDBMS. For example, \cite{icde:TranWRC09} assume that user-given keyword queries implicitly represent structured triple-pattern queries over RDF. They aim to find the top-$k$ structured queries that are relevant to a keyword query, which essentially extends the candidate network enumeration problem in RDBMS to RDF. \cite{cikm:ElbassuoniB11} and \cite{cikm:BicerTN11} study ranking models and algorithms for the results of those structured queries over RDF. \cite{tkde:LiLDK14} tries to find the top-$k$ entities that are reachable from all the keywords in the query over RDF. \cite{pvldb:WuYSIY13} proposes a new summarization language which improves result understanding and query refinement. It takes all the answers (subgraphs in RDF) to structured queries as input and output a summarization which is as concise as possible and satisfies certain coverage constraint.

\stitle{Searching aggregations in multidimensional data.}
A major motivation of our work is that a meaningful answer to a keyword query may be a collection of tuples/tuple joins, which need to be aggregated
before being output. This idea is also explored in multidimensional text data by \cite{tkde:DingZLHZSO11, kais:ZhouP12}. With a different data model and application scenarios, an answer there is a ``group-by'' on a subset of dimensions such that all keywords are contained in the aggregated tuples. In \cite{kais:ZhouP12}, how to enumerate all valid and minimal answers is studied, and in \cite{tkde:DingZLHZSO11}, scoring models for those answers and efficient algorithms to find the top-$k$ are proposed.
%

\section{Conclusions}
\label{sec:conclude}

We introduce the $d$-height tree pattern problem in a knowledge base for keyword search. Formal models of tree patterns are defined to aggregate subtrees in a knowledge graph which contain all keywords in a query.
Such tree patterns can be used to better understand the semantics of keyword queries and to compose table answers for users.
We propose path-based indexes and efficient algorithms to find tree patterns for a given keyword query. To further speed up query processing, a sampling-based approach is introduced to provide approximate top-$k$ with higher efficiency. Our approaches are evaluated using real-life datasets.


{\scriptsize
\bibliographystyle{abbrv}
\bibliography{ref}  
}


\newpage

\begin{appendix}

\section{Proof of Theorem 1}
\begin{proof}
It is easy to show that \countpat is in \sharpp, because for any tree pattern we can verify whether it is valid in polynomial time. To complete the proof, we need to prove its \sharpp-hardness by a reduction from the {$s$-$t$ {\sc Paths} problem}: counting the number of simple paths from node $s$ to $t$ in a directed graph $\graph = (\entities, \edges)$, which is proved to be \sharpp-Complete in \cite{siamcomp:Valiant79}.

For any instance of $s$-$t$ {\sc Paths} in a directed graph $\graph = (\entities, \edges)$, we first create a knowledge graph $\graph^2 = (\entities_2, \edges_2, \type, \attr)$ as follows: i) create two copies of the directed graph $\graph$, denoted by $(\entities', \edges')$ and $(\entities'', \edges'')$, and let $s'$/$s''$ and $t'$/$t''$ be the corresponding nodes of $s$ and $t$, respectively, in the two copies; ii) create a ``root'' node $r$ and two directed edges $(r,s')$ and $(r, s'')$; iii) let $\entities_2 = \entities' + \entities'' + \{r\}$ and $\edges_2 = \edges' + \edges'' + \{(r,s'), (r,s'')\}$; and iv) let types $\type$ on the nodes and attributes $\attr$ on the edges be unique, and text descriptions on nodes/edges (types) be unique. Second, let $\query$ be a keyword query with the two words from the text in entities corresponding to $t'$ and $t''$. We can show that the answer to the $s$-$t$ {\sc Paths} instance with input $\graph$ is $N$ iff the answer to the \countpat instance with input $\graph^2$, $\query$, and $d = |\entities|+1$ is $N^2$. So the proof is completed.
\end{proof}

\section{Proof of Theorem 5}
\begin{proof}
For pattern $P_i$ ($i=1$ or $2$), from the definition, we can decompose its score $\scoreexact_i=\score(P_i, \query)$ among all candidate roots:
\begin{align}
\scoreexact_i & = \score(P_i, \query) = \!\!\!\!\!\!\!\! \sum_{T \in \trees(P_i)} \!\!\!\!\!\! \score(T, \query) = \sum_{r \in \entities} \sum_{T \in \trees_r(P_i)} \!\!\!\!\!\!  \score(T,q) \nonumber
\\
& = \sum_{r \in \entities} \scoreexact_i(r), \label{equ:proof:scoredecompose}
\end{align}
where $\trees_r(P_i)$ is the set of valid subtrees with pattern $P_i$ and rooted at node $r$, and $\scoreexact_i(r) = \sum_{T \in \trees_r(P_i)} \score(T,q)$ is the sum of relevance scores of all valid subtrees rooted at $r$ for pattern $P_i$. When $\trees_r(P_i) = \emptyset$, define $\scoreexact_i(r) = 0$. We suppose $\scoreexact_1 > \scoreexact_2$.

When \algLinearEnumTopK runs with parameter $\Lambda = 0$, we get a random subset of candidate roots $R^+ \subseteq \entities$ across different root types, such that each candidate root is selected into $R^+$ with probability $\rho$ (line~8). Then we can estimate $\score(P_i, \query)$ as:
\[
\scoreapprox(P_i, \query) = \frac{1}{\rho} \sum_{r \in R^+} \scoreexact_i(r).
\]

It is not hard to show that $\ep{\scoreapprox(P_i, \query)} = \score(P_i, \query) = \scoreexact_i$.

Now define $|\entities|$ independent random variables:
\[
X(r) = \left\{ 
  \begin{array}{l l}
    \scoreexact_1(r) - \scoreexact_2(r) & \quad \text{with probability $\rho$};\\
    0 & \quad \text{with probability $1-\rho$}.
  \end{array} \right.
\]

From the definitions, we have
\begin{align}
\pr{\scoreapprox(P_1, \query) < \scoreapprox(P_2, \query)} & = \pr{\sum_{r \in R^+} \scoreexact_1(r) < \sum_{r \in R^+} \scoreexact_2(r)} \nonumber
\\ 
& = \pr{\sum_{r \in R^+} (\scoreexact_1(r) - \scoreexact_2(r)) < 0} \nonumber
\\
& = \pr{\sum_{r \in \entities} X(r) < 0}. \label{equ:proof:precision:eq}
\end{align}

From the linearity of expectation and \cequ\eqref{equ:proof:scoredecompose}, we can show that $\ep{\sum_{r \in \entities} X(r)} = (\scoreexact_1 - \scoreexact_2) \cdot \rho$. So we have
\begin{align}
& \pr{\sum_{r \in \entities} X(r) < 0} \nonumber
\\
= & \pr{\sum_{r \in \entities} X(r) - \ep{\sum_{r \in \entities} X(r)} < -(\scoreexact_1 - \scoreexact_2) \cdot \rho} \nonumber
\\
\leq & \exp\left(-\frac{2((\scoreexact_1 - \scoreexact_2) \cdot \rho)^2}{\sum_{r \in \entities} (\scoreexact_1(r) - \scoreexact_2(r))^2}\right) \label{equ:proof:precision:hoeffding}
\\
\leq & \exp\left(-\frac{2((\scoreexact_1 - \scoreexact_2) \cdot \rho)^2}{\left(\sum_{r \in \entities} (\scoreexact_1(r) + \scoreexact_2(r))\right)^2}\right) \label{equ:proof:precision:sumineq}
\\
= & \exp\left(-2\left(\frac{\scoreexact_1 - \scoreexact_2}{\scoreexact_1 + \scoreexact_2}\right)^2 \cdot \rho^2 \right). \label{equ:proof:precision:bound}
\end{align}

\eqref{equ:proof:precision:sumineq} is because of the inequality: $\sum_i(x_i - y_i)^2 \leq (\sum_i x_i + \sum_i y_i)^2$ for $x_i, y_i \geq 0$. \eqref{equ:proof:precision:bound} is directly from \cequ\eqref{equ:proof:scoredecompose}. And \eqref{equ:proof:precision:hoeffding} is from Hoeffding's inequality in the following lemma where we have each independent random variable $X(r)$ bounded between $\scoreexact_1(r) - \scoreexact_2(r)$ and 0, and set $t = (\scoreexact_1 - \scoreexact_2) \cdot \rho$.

\begin{lemma} {\bf (Hoeffding's Inequality \cite{book:2009DubhashiP})} Let $X_1, X_2, \ldots, X_n$ be independent bounded random variables such that $X_i \in [a_i, b_i]$ with probability 1. Then for any $t > 0$, we have
\[
\pr{\sum_{i=1}^n X_i - \ep{\sum_{i=1}^n X_i} \leq -t} \leq \exp\left(-\frac{2t^2}{\sum_{i=1}^n(b_i - a_i)^2}\right).
\]
\end{lemma}

Putting \eqref{equ:proof:precision:eq} and \eqref{equ:proof:precision:bound} together, the proof is completed.
%
\end{proof}

\begin{figure}[t]
\center
\includegraphics[width=0.48\textwidth]{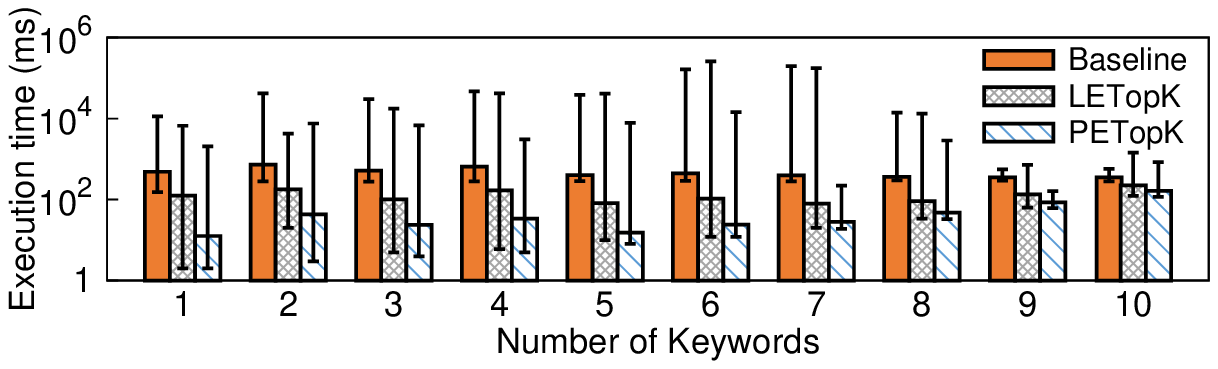}
\vspace{-1.8em}
\caption{Execution time for queries with different numbers of keywords in Wiki dataset}
\vspace{-1.6em}
\label{fig:time_vs_keywords}
\end{figure}

\section{Additional Experiments}

\stitle{Exp-A-I: Varying number of keywords.}
The performance of our algorithms is not sensitive to the number of keywords in a query. In Wiki dataset, we evaluate the 500 queries, in which the number of keywords vary from 1 to 10. We plot the min / average / max execution time of our algorithms in \wiki for different numbers of keywords in \cfig\ref{fig:time_vs_keywords}. We find that the performance of our algorithms does not deteriorate for more keywords (sometimes they are even faster). The reason is as follows: while the time complexity of both \algPE and \algLE increases as the number of keywords increases, the real bottleneck is the number of valid subtrees. For \algPE, with more keywords, line~5 of \calg\ref{alg:patternenum} is more likely to generate less number of candidate roots, and thus line~7 generates less number of valid subtrees. For \algLE, as can be seen in \cthm\ref{thm:linearenum:correct}, its complexity is linear in the number of keywords ($m$) but the dominating factor is the number of valid subtrees ($N$).

\remove{
\newpage
\subsection*{Improvement Suggested by Reviewer II}

{\sf PETopK-r1}\xspace incorporates the first optimization by reviewer one, and {\sf PETopK-r2}\xspace incorporates the first and second optimizations by reviewer one.

\begin{figure}[htbp]
\includegraphics[width=0.48\textwidth]{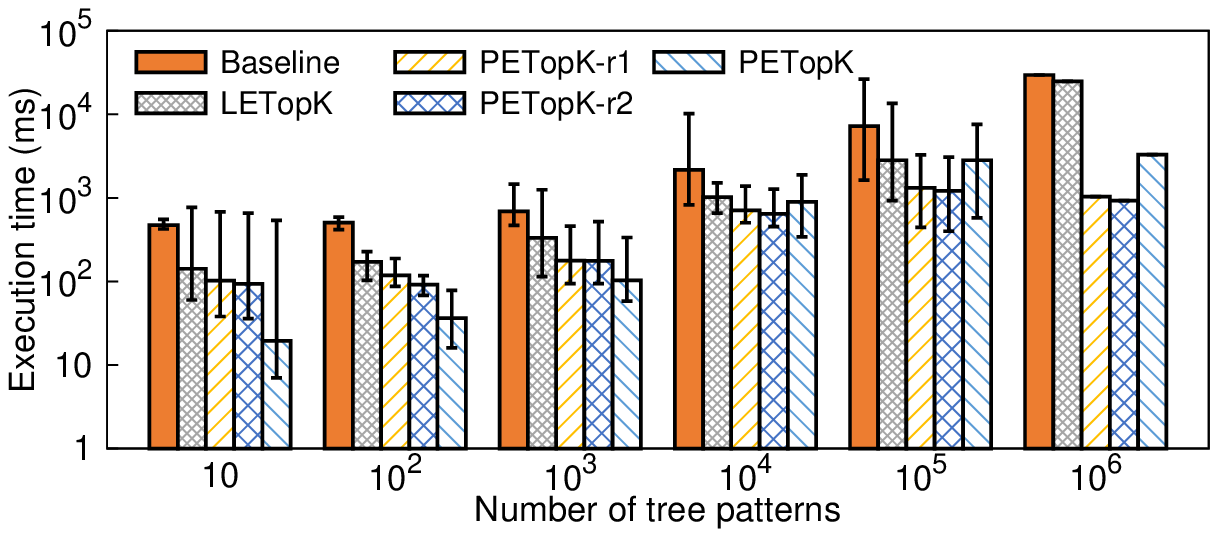}
\vspace{-2em}
\caption{Execution time}
\vspace{-1em}
\end{figure}

\begin{table}[htbp]
\begin{tabular}{l||lll}
\hline
& {\sf PETopK-r1} & {\sf PETopK-r2} & \algPE\\ \hline \hline
38 wiki queries & 0.307 & 0.282 & 0.206 \\
3 large wiki queries & 46.743 & 43.392 & 92.826 \\ \hline
\end{tabular}
\vspace{-1em}
\caption{Average execution time overall and for the three largest queries}
\vspace{-1em}
\end{table}
}
\end{appendix}

\end{document}